\documentclass[useAMS,usenatbib,babel]{mn2e}
\usepackage{mathpazo}

\usepackage[english,english]{babel}
\usepackage{amsmath}
\usepackage{amssymb,amsfonts,textcomp}
\usepackage{array}
\usepackage{supertabular}
\usepackage{hhline}
\usepackage{hyperref}
\usepackage[usenames]{color}
\hypersetup{dvips, colorlinks=true, linkcolor=blue, citecolor=blue, filecolor=blue, urlcolor=blue}
\usepackage[dvips]{graphicx}

\def\gtrsim{\lower.5ex\hbox{$\; \buildrel > \over \sim \;$}}
\usepackage{graphicx}

\definecolor{grey}{rgb}{0.75,0.75,0.75}
\definecolor{Orange}{rgb}{1.0,0.5,0.15}
\definecolor{brown}{rgb}{0.7,0.25,0.0}
\definecolor{pink}{rgb}{1.0,0.5,0.5}
\definecolor{darkerred}{rgb}{0.8,0,0}
\definecolor{darkerblue}{rgb}{0,0,0.8}
\definecolor{Blue}{rgb}{0,0.08,0.65}
\definecolor{Red}{rgb}{0.65,0.08,0.05}
\definecolor{Green}{rgb}{0.15,0.45,0.25}

\begin{document}

\author[ Y. Dubois et al. ]{
\parbox[t]{\textwidth}{
 Yohan Dubois$^{1,2}$\thanks{E-mail: dubois@iap.fr}, Christophe Pichon$^{1}$, Martin  Haehnelt$^{3}$, Taysun  Kimm$^{2}$, \\
 Adrianne~Slyz$^{2}$, Julien Devriendt$^{2,4}$, and  Dmitry Pogosyan$^{5}$ }
\vspace*{6pt} \\
$^{1}$ Institut dÕAstrophysique de Paris, UMR 7095, CNRS, UPMC Univ. Paris VI, 98 bis boulevard Arago, 75014, Paris, France\\
$^{2}$ Sub-department of Astrophysics, University of Oxford, Keble Road, OX1 3RH, Oxford, UK.\\
$^{3}$ Institute of Astronomy and Kavli Institute for Cosmology, Madingley Road, CB3 0HA, Cambridge, United Kingdom.\\
$^{4}$ Observatoire de Lyon, UMR 5574, 9 avenue Charles Andr\'e, F-69561, Saint Genis Laval, France. \\
$^{5}$ Department of Physics, University of Alberta, 11322-89 Avenue, Edmonton, T6G 2G7, Alberta, Canada.
}
\date{Accepted 2012 April 21.  Received 2012 April 16; in original form 2011 December 11}

\title[Feeding supermassive black holes with cosmic gas]{
Feeding compact bulges and supermassive black holes with low angular-momentum cosmic gas at high redshift}

\maketitle

\begin{abstract}
{We use cosmological hydrodynamical simulations to show that
a significant  fraction of the gas  in high redshift rare massive halos falls nearly radially 
to their very centre on extremely short timescales.  This process results in the formation of very compact bulges 
with  specific angular momentum a  factor $5-30$
smaller than the average  angular momentum of the baryons  in the whole
halo. Such low angular momentum originates both from segregation and effective
cancellation when the gas flows to the centre of the
halo  along  well defined cold filamentary streams. These filaments penetrate
deep inside the halo and connect to the bulge from  multiple 
rapidly changing  directions.  Structures  
falling in  along  the filaments (satellite galaxies) or formed  
by gravitational instabilities triggered by the inflow (star clusters) further
reduce the  angular momentum of the gas in the bulge.
Finally, the fraction of gas radially falling to the centre  appears to
increase  with the mass of the halo; we argue that this is most
likely due to an enhanced cancellation of angular momentum in rarer
halos which are  fed by more isotropically distributed
cold streams. Such an increasingly efficient funneling of low-angular
momentum gas to the centre of very massive 
halos at high redshift may account for the rapid pace at which the most massive 
super massive black holes grow to reach observed masses around $10^9$M$_\odot$ 
at an epoch when the Universe is barely 1 Gyr old.}
\end{abstract}


\section{Introduction}

Supermassive black holes (BH) have been established to be ubiquitous  at the centre of local
galactic bulges and their mass has been shown to correlate well with the stellar mass and perhaps 
even more strongly  with the stellar velocity dispersion of these bulges
\citep{magorrianetal98,  tremaineetal02, haring&rix04}. 
As the accretion of gas onto BHs can drive strong feedback from Active Galactic Nuclei (AGN) through spherical winds or collimated jets, 
and can potentially self-regulate the growth of the BH along with the cold baryon content of galaxies \citep{silk&rees98, haehneltetal98, king03}
AGN feedback is a natural candidate to explain the observed correlations. Indeed, this picture has been further substantiated by several numerical implementations 
of such feedback using either semi-analytical models \citep{boweretal06, crotonetal06, somervilleetal08} or hydrodynamical simulations \citep{dimatteoetal05, booth&schaye09, duboisetal12}.

The issue is further complicated by the fact that the most supermassive BHs (several $10^9\, \rm M_\odot$) seem to be
already in place  at  $z\approx 6-7$,  i.e. less than a Gyr  after the
Big-Bang \citep{willottetal03, fanetal06,jiangetal09,mortlocketal11}. 
Growth to such large masses in such a short time scale
constitutes  a significant challenge for any model of BH evolution, especially those with strong AGN feedback, 
as it requires sustained feeding at close to the Eddington  accretion
rate \citep{haiman04, sijackietal09}.  
However, considering that a substantial fraction  of the sky had
to be surveyed to discover the luminous QSOs powered by these very
massive high-redshift BHs ($10^{-9} \, \rm Mpc^{-3}$ comoving number density of high-redshift quasars according to~\citealp{fanetal06}), it appears likely that the halos
hosting these BHs  are the most  massive formed  at these
early times and thus are very rare objects. Obviously, the main caveat of this 'rareness' argument is that since these young BHs are thought to be growing fast because of the presence of large amounts of gas in their surroundings, a larger number of their quasar counterparts could possibly be obscured in optical wavebands \citep{alexanderetal03, treisteretal11, willott11} and only be visible  in X-rays~\citep[e.g.][]{daddietal07} and the Infrared. This selection effect could increase the {\it true} number of very bright quasars/super massive BHs at high redshift considerably. 

Nevertheless, even when adopting the view that these objects are rare, one still needs to propose an effective mechanism  
to funnel low angular momentum gas all the way down to the sphere of influence of the BHs at the very 
centre of  their very massive halos hosts.
Hydrodynamical simulations suggest that high-redshift quasars are accreting gas at rates close to their Eddington (or super-Eddington) limit 
because large amounts of cold gas are indeed trapped in the centre of galaxies, and that the fraction of Eddington-limited growth of BHs diminishes with time
as this gas reservoir is depleted~\citep{sijackietal07, dimatteoetal08, duboisetal12}.   \cite{dimatteoetal12} have
emphasized  that filamentary infall of cold gas is the major mode of
gas  supply for the continuous feeding of BHs similar to what is discussed
for the build-up of galaxies at intermediate redshift \citep{keresetal05, agertzetal09}.
Note that this is  a very different mode of gas supply from that due to 
the often invoked more episodic feeding of BHs  due to large amounts of gas 
driven to  galactic nuclei by  (major) mergers \citep{mihos&hernquist96, kauffmann&haehnelt00, hopkinsetal06, mayeretal10}

\cite{pichonetal11} and  \cite{kimmetal11}  have recently used hydro cosmological Adaptive Mesh Refinement (AMR)  simulations  to investigate  the angular
momentum properties and the infall of gas in Milky Way class halos at low 
and intermediate redshifts with a view to study the formation and evolution 
of their central galaxy discs. With a similar technique~\cite{bournaudetal11} have performed
simulations of isolated galaxy discs with parsec resolution to  show
that Toomre instabilities drive strong inflows of gas to the centre
of these discs.  

We build here on this work, by investigating the more massive and rarer  
halos expected  to host the most massive supermassive BHs at  high redshift.
We therefore concentrate our exploratory study on the gas with the lowest angular
momentum falling to the centre of these rare, massive  halos. 

\begin{figure*}
   \includegraphics[width=\columnwidth]{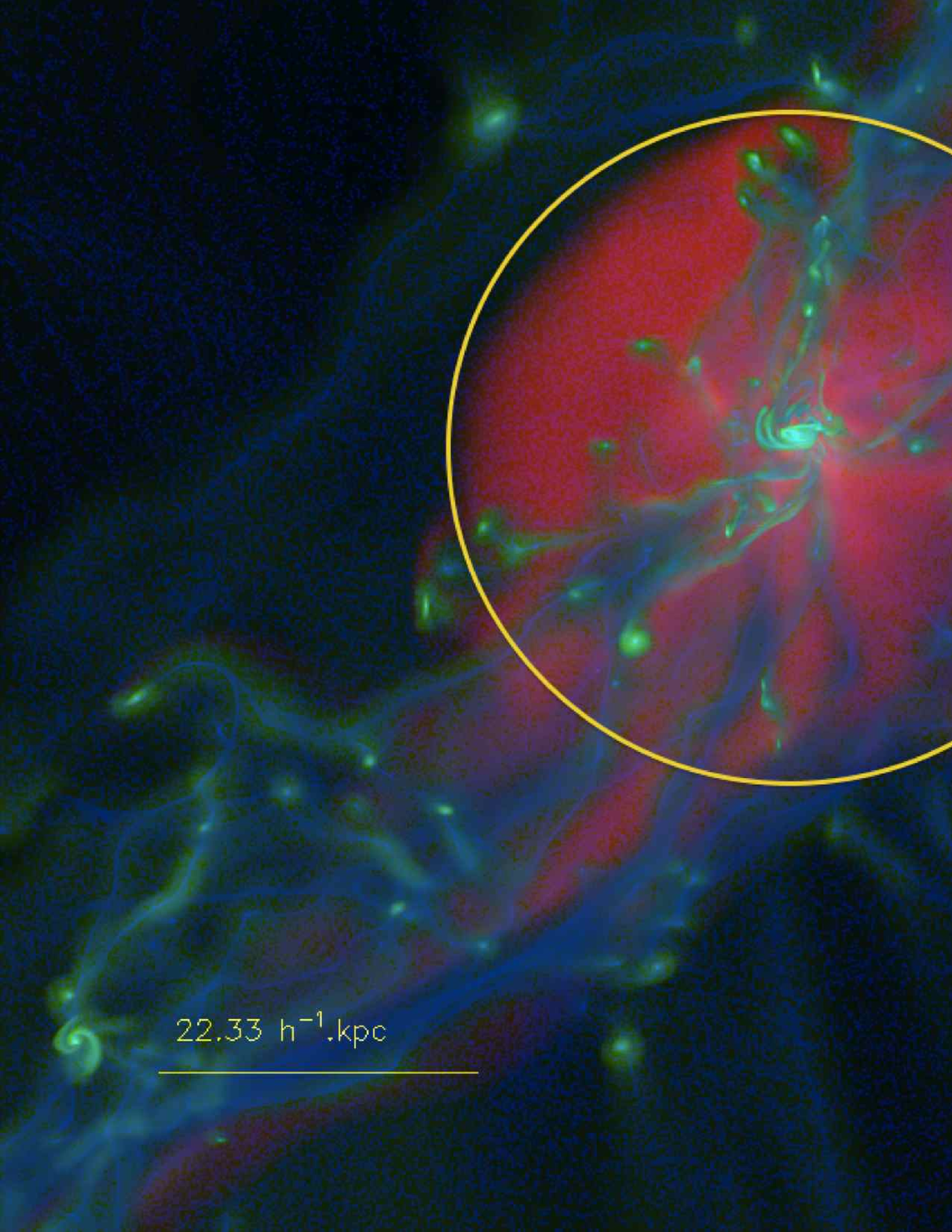}
   \includegraphics[width=\columnwidth]{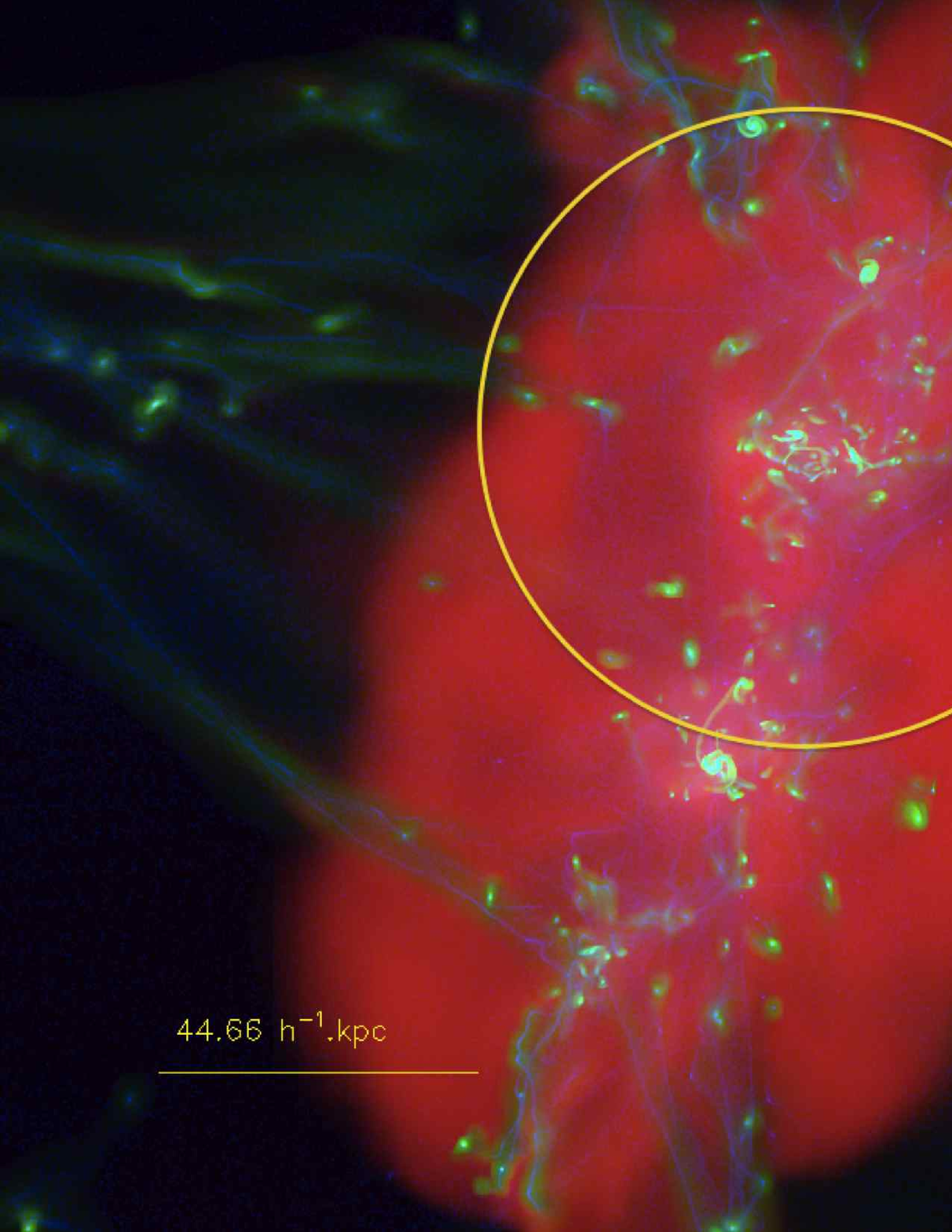}
   \includegraphics[width=\columnwidth]{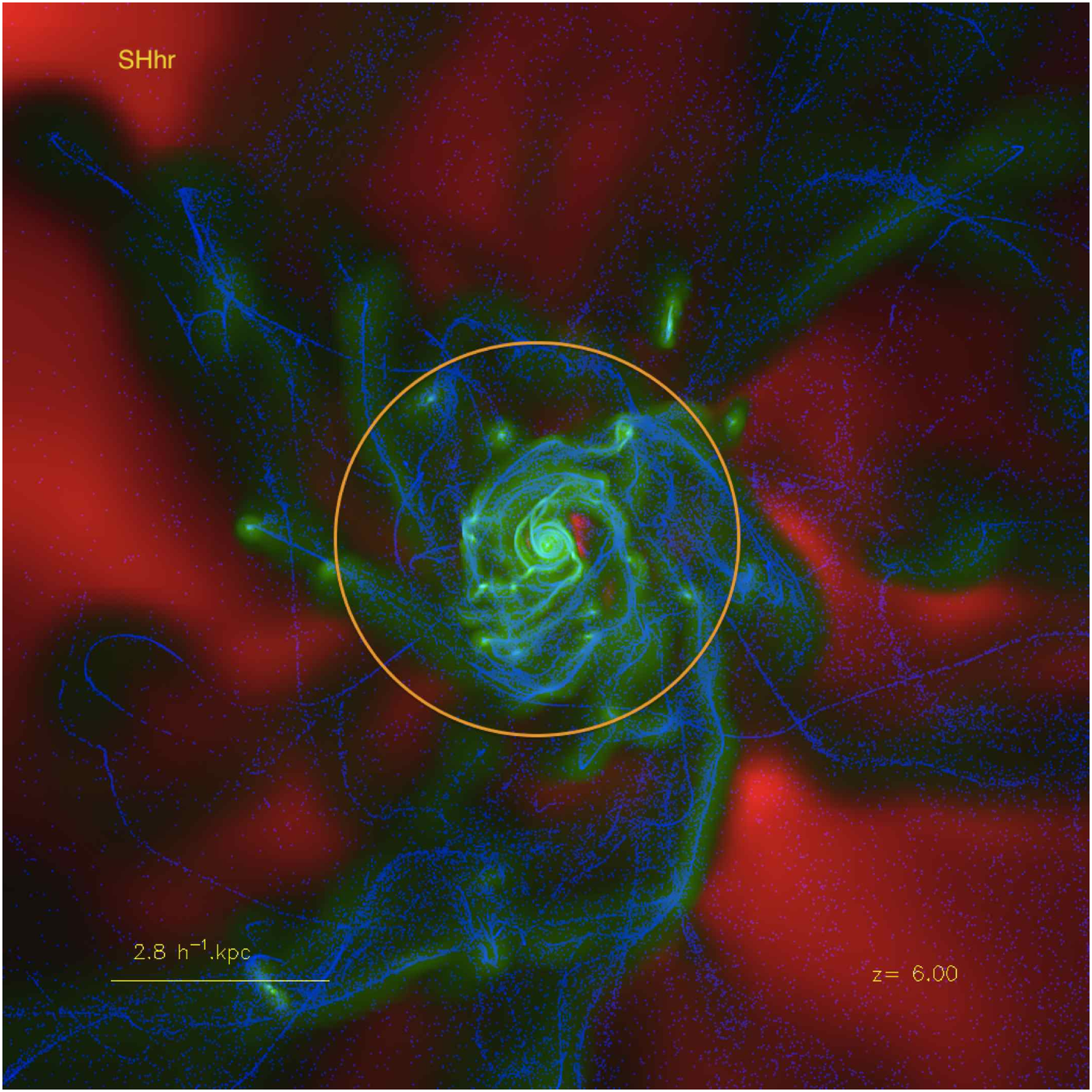}
   \includegraphics[width=\columnwidth]{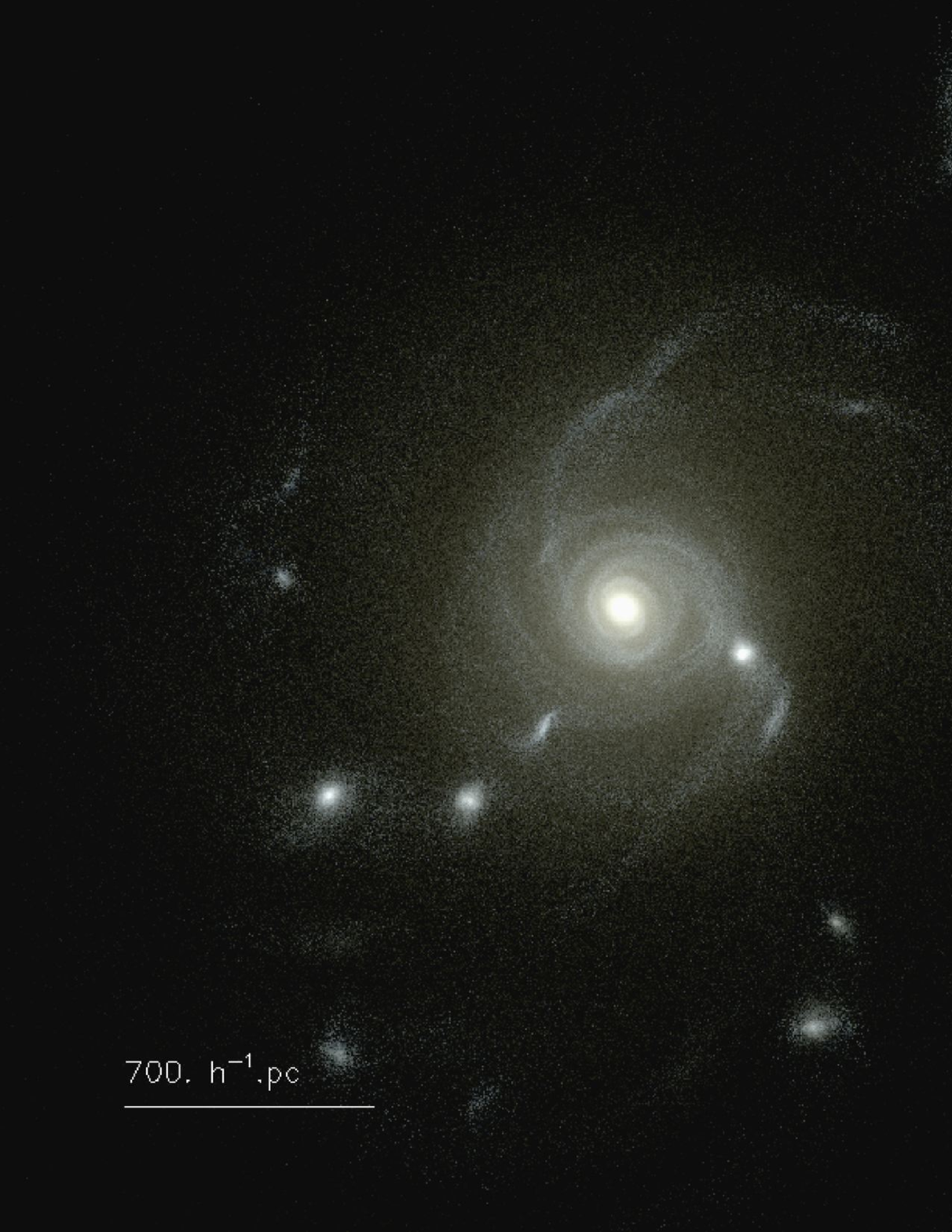}
      \caption{{\sl Top panels:} 
Polychromatic view of the two halos in the SH and the LH simulation at $z=6$. 
Gas density is color coded in green, gas temperature in red, 
and tracer particles in blue. The circles in the two top panels corresponds to  the virial radius $r_{\rm vir}$.
The {\sl bottom left panel} displays the inner region 
of the top left panel simulated with higher resolution (SHhr),
showing the transient  disc and its connected filaments as discussed in the text. 
The  circle corresponds to 0.1 $r_{\rm vir}$ (see table~\ref{tab:feature} for numerical values).
The {\sl bottom right panel } displays the stellar emission in the inner region of the bottom
left panel  as it would be observed  in  ugr filter bands. Note the 
compact bulge (the saturated spherical region at the centre) and the significant number of satellites  flowing in along the
cold streams  on their way to merge with the central galaxy. 
Finally, we point out that even though in projection the hot gas phase seems to dominate, this halo is still 
mainly accreting cold gas.
 }
\label{fig:visual}
\end{figure*}

 Whilst \cite{dimatteoetal12} have recently investigated  this problem using a
simulation with a large enough volume ($533\, h^{-1}\cdot\rm Mpc$) to capture halos
sufficiently  massive to plausibly host these $\sim 10^9\, \rm M_\odot$ BHs,
 we take a different approach here and  focus our study on two
individual very massive halos at high redshift which we follow
using a high-resolution cosmological hydrodynamics re-simulation technique.
 The main aim here is to better resolve the
dynamics and angular momentum history of the inflowing cold gas as it
falls to the centre of halos, forms a galactic bulge and potentially
feeds the growth of the central supermassive BH. 

\begin{figure*}
   \includegraphics[width=\columnwidth]{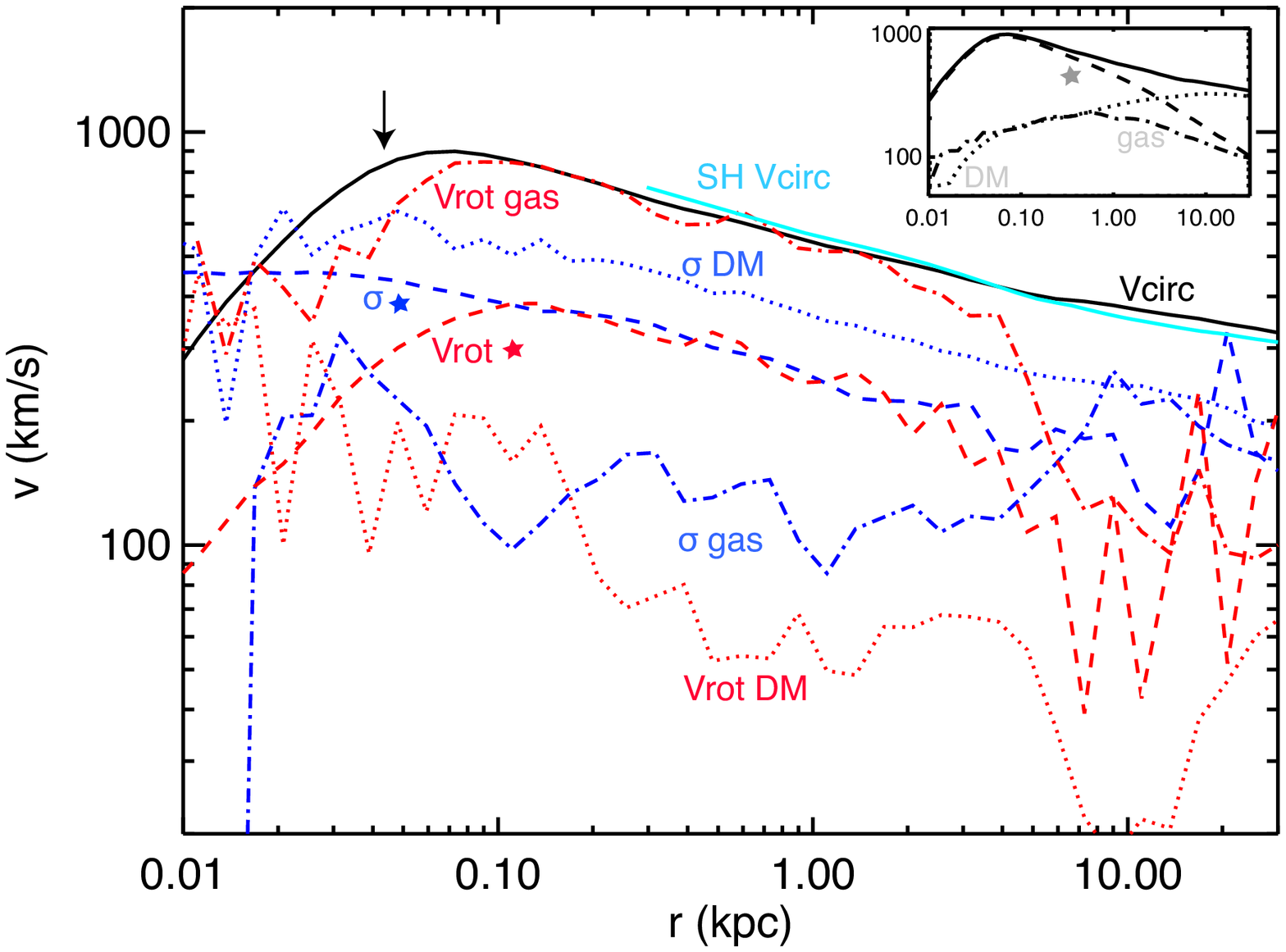}   
   \includegraphics[width=\columnwidth]{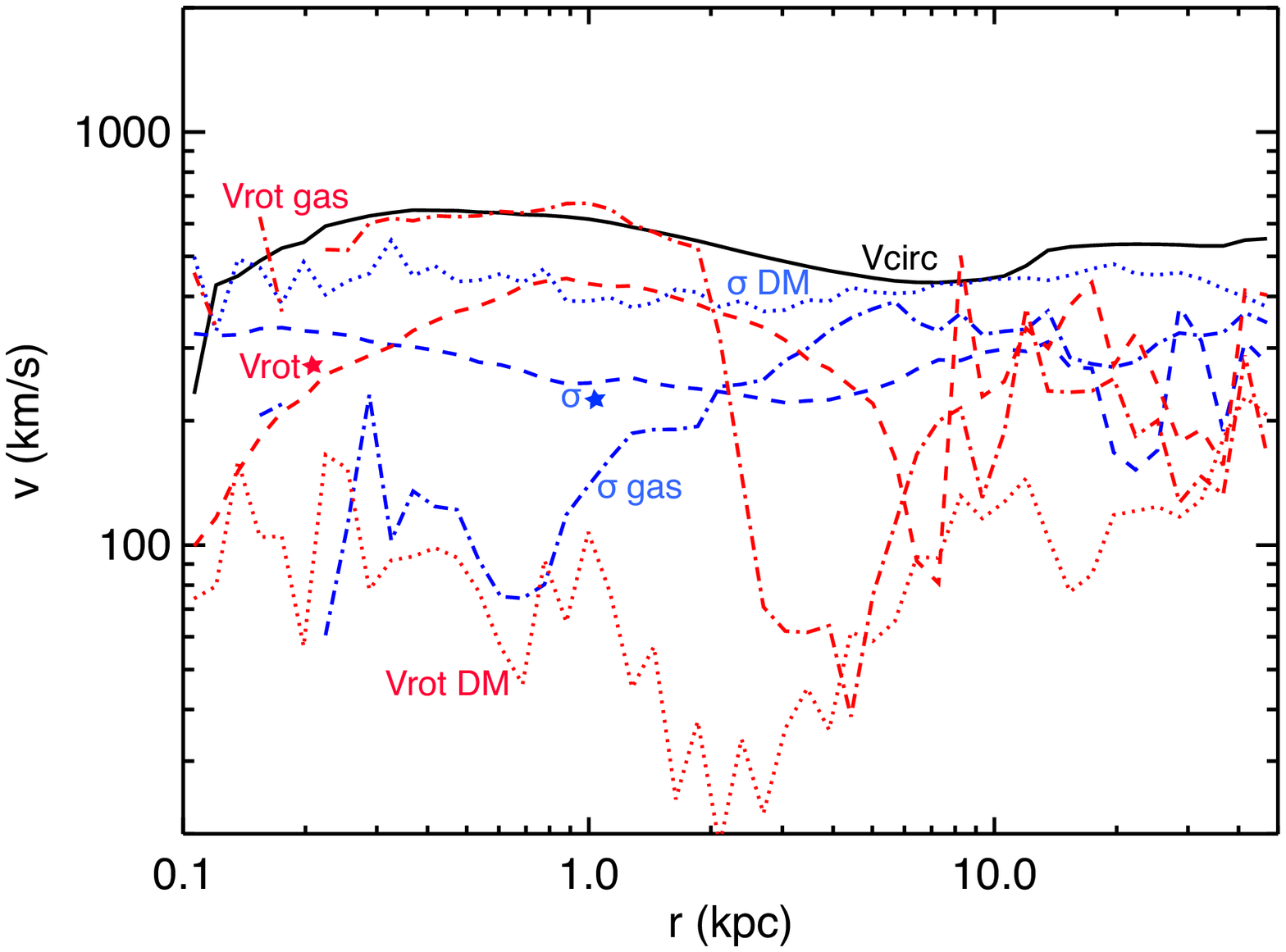}   
       \caption{Circular velocities (black lines and inset), radial velocity dispersion (blue lines) and 
         rotational velocities (red lines) of the DM (dotted
         lines), the stars (dashed lines), and the gas (dot-dashed
         lines) for the SHhr simulation (left panel), and the
         LH simulation (right panel) at $z=6$. The blue line corresponds to the SH run.
         The bulge radius for the SHhr simulation is shown as a vertical arrow.
         At this  redshift, these galaxies have the rotation curve of  very compact ellipticals.
   }
\label{fig:vrot_vcirc}
\end{figure*}

The paper is organized as follows. Section~\ref{section:numerics}
describes the numerical setup of our simulations.
Section~\ref{sec:mass} investigates the angular momentum evolution 
and the typical trajectories of the baryonic material forming the bulge. 
Section~\ref{sec:discussion} discusses the 
implications for BH build up at high redshift
while Section~\ref{sec:conclusion}  gives our conclusions.

\section{ The Numerical Simulations}
\label{section:numerics}

\begin{figure*}
	         \includegraphics[width=2.1\columnwidth]{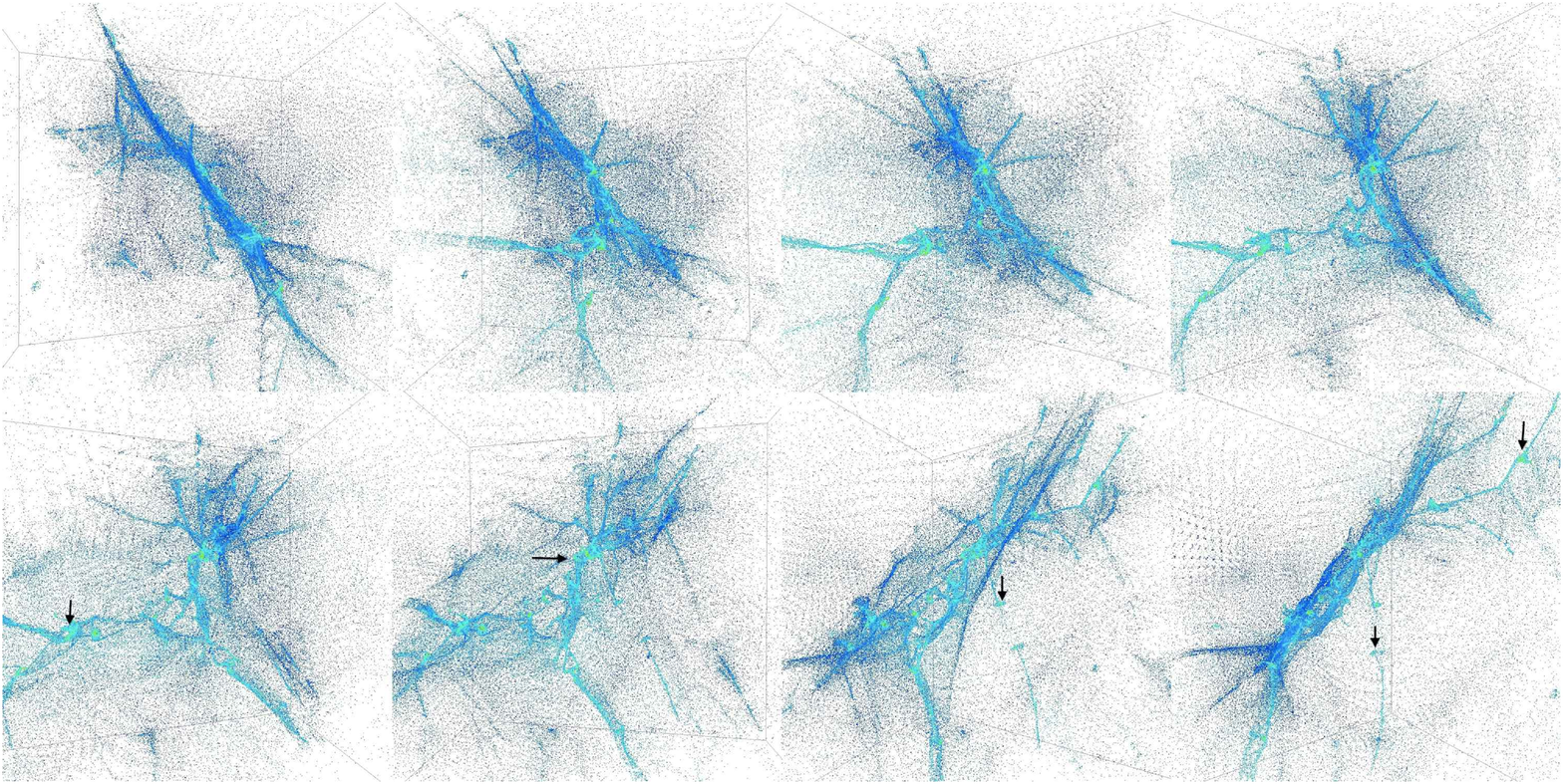}  	
  \caption{ Distribution of tracer particles at $z=9$  
    identified as belonging to the bulge of the SH halo at $z=6$.
{\sl  From top to bottom and left to right}, the observer is
spiraling in towards the center of the web in the anti-clockwise direction around a
vertical axis. The color coding represents the  local log density of tracer particles 
(from blue: low density to yellow:high density).
 The size of the box is approximately 100 kpc  across on the top left panel.
The web-like filamentary structure of the gas
distribution which ends up in the bulge is rather complex, though one filament embedded
in a main wall dominates  (best seen edge-on on the top left and bottom right panels). Note the disc-like
features (marked by vertical arrows) which have formed perpendicular to the filaments. The forming bulge is marked by an horizontal arrow. We provide an animation 
which allows to better see this at {\em\tt
http://www.iap.fr/users/pichon/BH/}. 
}
\label{fig:visual-tracer}
\end{figure*}

\begin{figure*}
   \includegraphics[width=8.75cm]{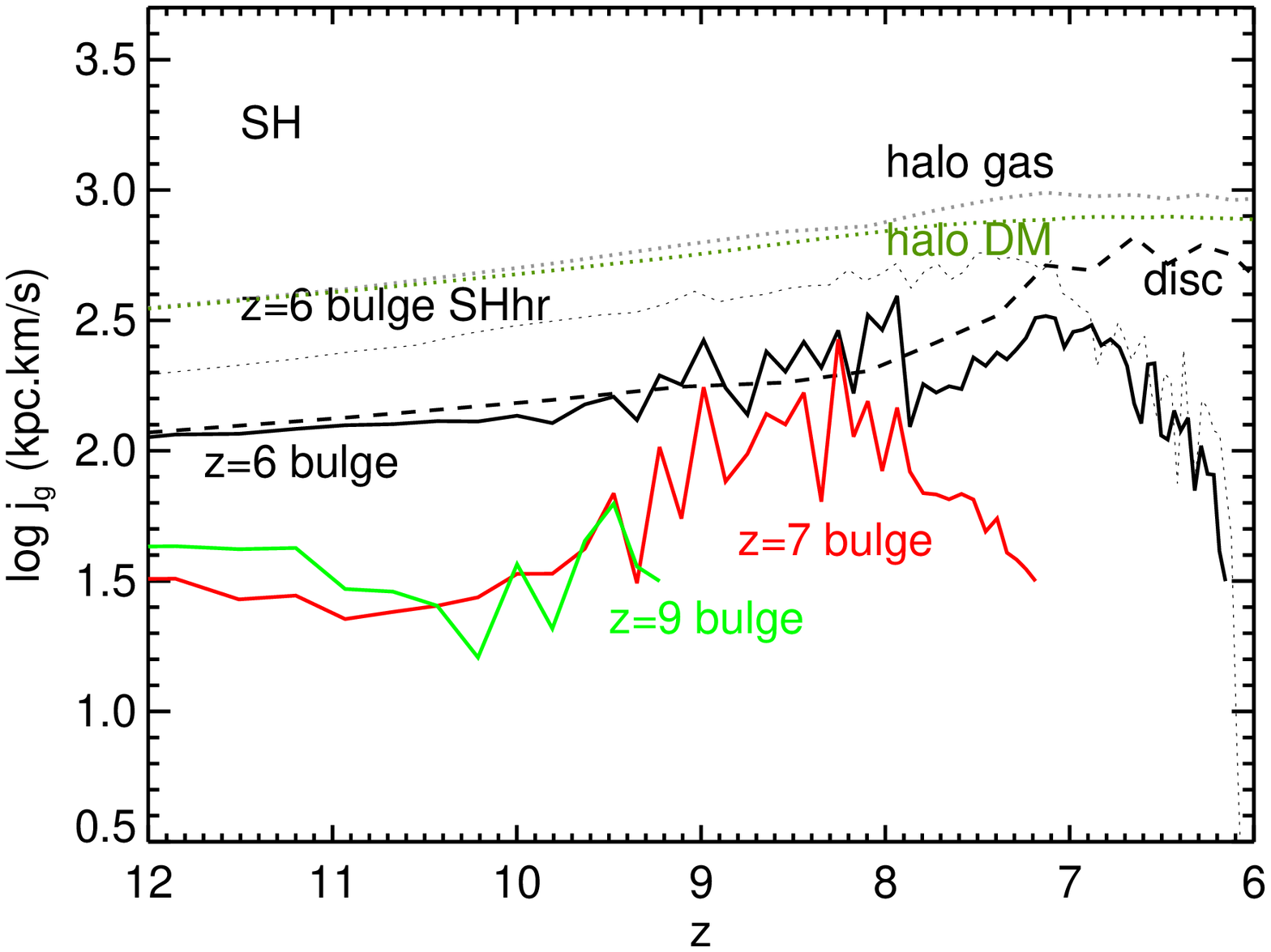}
   \includegraphics[width=8.75cm]{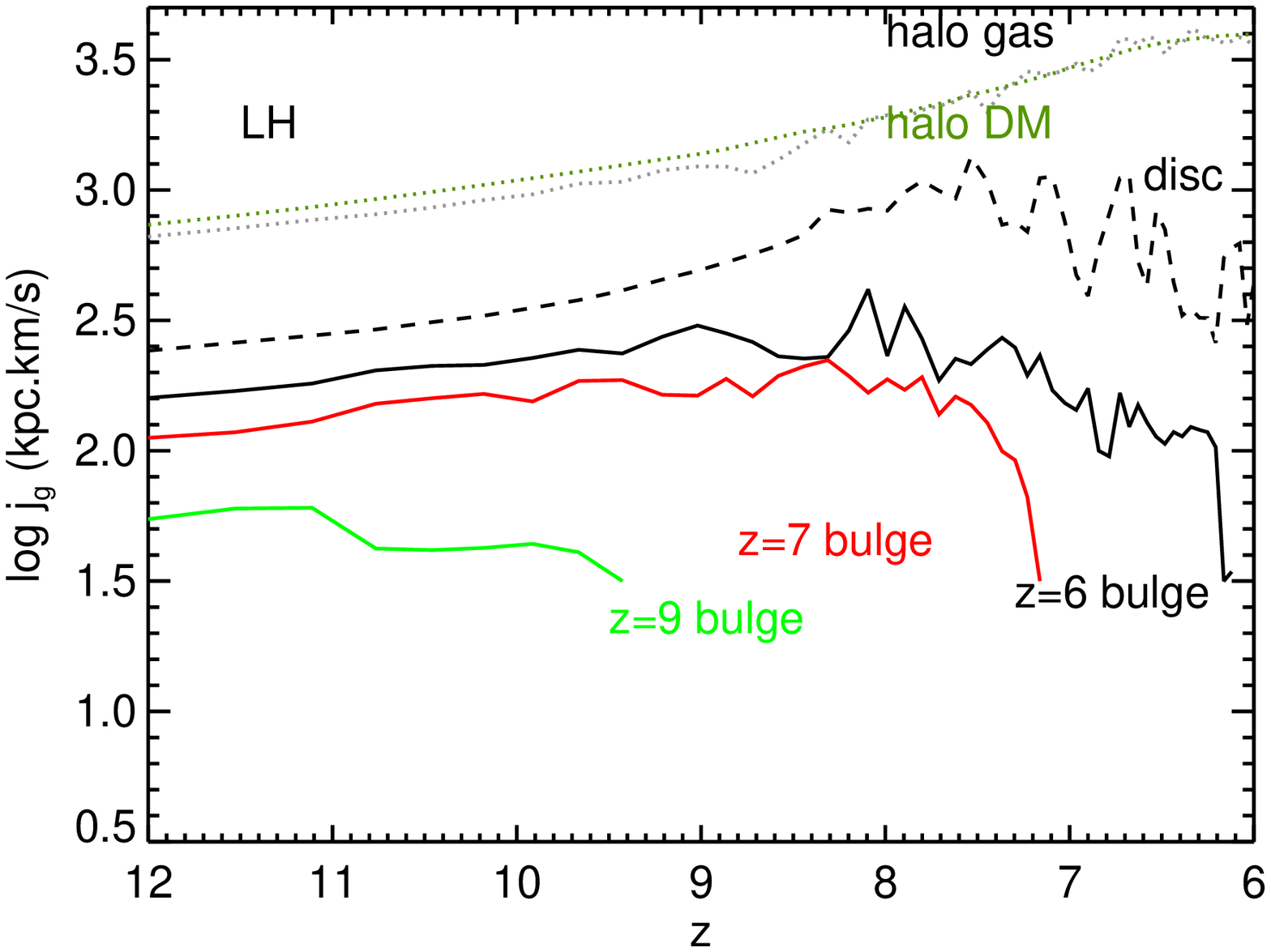}
      \caption{ \emph{Left panel:} Evolution of the specific angular momentum of the
gas traced by the tracer particles identified to be part of the disc at  $ 0.2 < r < 2 \, \rm kpc $ in the SH simulation (dashed line), for tracer particles identified to be part of the bulge at $r \le
0.2 \, \rm kpc$ (solid lines) (or at $r \le 0.04 \, \rm kpc$ for the SHhr simulation, black dotted curve), and for tracer particles within $r_{\rm rvir}$ (grey dotted line) or DM particles (green dotted line) within $r_{\rm rvir}$ at $z=6$. The bulge tracer particles are identified as  falling into the bulge at $z=6$ (black solid curve for the SH simulation, black dotted curve for the SHhr simulation), $z=7$ (red curve),
and $z=9$ (green curve). The specific angular momentum is  computed
relative to the centre of mass of the ensemble of tracer particles (bulge, disc or halo) at a given time.
    \emph{Right panel: } same as left panel for the LH simulation. 
The prominent drop in angular momentum is due to the cancellation of angular
momentum of material streaming into the bulge from different directions.
The larger drop in  the SHhr simulation  is due to the 
more efficient migration of clumps in the higher resolution 
simulation.
}
\label{fig:momemtumvsz}
\end{figure*}

We have performed two simulations with two different constrained
initial conditions, generated using the {\tt constrfield} package of
{\tt mpgrafic} \citep{prunetetal08}. The  {\tt constrfield} package
generates realizations of the initial density field, subject 
to constraints \citep{BBKS,hoffmanribak} 
on the linear properties of an arbitrary set of spherical patches.
A patch is defined by its position, size (or mass) and the window function.
Any of the following patch properties, averaged over the window,
can be specified as the constraints: the overdensity, the velocity, 
the velocity shear, the gradient of the density (set to zero if a patch
represents the peak-patch progenitor of the collapsing halo) 
and the matrix of the  second derivatives of the density that defines the initial
ellipticity of the density profile in a patch.
The {\tt mpgrafic} realization of {\tt constrfield} follows
closely the code developed for Cosmic Web studies in \cite{bkp96}.

The first simulation was chosen to produce a halo with a Virial mass of $M_{\rm vir}=2 \cdot 10^{15} M_{\odot}$ at $z=0$ (which has a mass of $M_{\rm vir}=5 \cdot 10^{11} M_{\odot}$
at $z=6$)  and the second a halo with a mass of $M_{\rm vir}=2.5 \cdot 10^{12}
M_{\odot}$ at $z=6$.  In the following, we  will refer to the two simulation  as SH and
LH, respectively. For our chosen cosmological parameters (see below) 
the entire Universe should contain about 
$\sim 2\cdot10^4$ halos as massive as the one in the SH simulation while we should find only one  halo as massive as the one in the LH simulation (but note that this depends very sensitively on the
assumed matter power spectrum normalization). Our simulations should therefore
represent  a rare and a very rare massive halo respectively, and we
refer to  Table~\ref{tab:feature} for the basic properties of the two halos.
 Note that while the halo in the SH simulation has a fairly typical spin parameter of $\lambda =0.05$ at $z=6$ the halo in LH
has a value twice higher, due to an ongoing merger. Such factor of two fluctuations
of the spin parameter  are not unusual during the build up of halos. 
To set up initial conditions, only the overdensity and peak (zero gradient)
constraints were used with a top-hat window filtering that corresponds to the chosen masses.
Overdensity values were selected to ensure that the halo collapses by the required redshift.

\begin{table}
\caption{Main properties of  the SH, SHhr and LH simulations at $z=6$. Stellar
bulge versus disc masses  are obtained with a bulge-disc
decomposition of the stellar mass profiles. Gas and DM masses in the bulge and
in the galaxy are measured at the radius where the stellar mass {\it effectively} reaches the mass inferred from 
the bulge/disc decomposition. }
\label{tabnames}
\center
\begin{tabular}{@{}|l|c|c|c|}
  \hline
 & SH & SHhr & LH \\
  \hline
  \hline
 $M_{\rm vir} (\rm M_{\odot})$ & $5\cdot10^{11}$ & $5\cdot10^{11}$ & $2.5\cdot10^{12}$   \\
  \hline
 $r_{\rm vir} (\rm kpc)$ & 36 & 36 & 60   \\
  \hline
 $V_{\rm vir} (\rm km/s)$ & 245 & 245 & 425   \\
  \hline
  \hline
 $\Delta x (\rm pc)$ & 135 & 17 & 135   \\
  \hline
 $M_{\rm res} (\rm M_{\odot})$ & $1.3\cdot10^6$ & $1.3\cdot10^6$ & $1\cdot10^7$   \\
  \hline
  \hline
$M^{\rm bulge}_{\rm star}(\rm M_{\odot})$ & $2.2 \cdot10^{10}$ &  $1.9 \cdot10^{10}$ & $3.6 \cdot10^{10}$ \\
  \hline
$M^{\rm disc}_{\rm star}(\rm M_{\odot})$ &$1.5 \cdot10^{10}$ & $2.9 \cdot10^{10}$ & $2.0 \cdot10^{10}$ \\
  \hline
$M^{\rm disc+bulge}_{\rm star}(\rm M_{\odot})$ & $3.9 \cdot10^{10}$ & $5.6 \cdot10^{10}$ & $5.8 \cdot10^{10}$ \\
  \hline
  \hline
$M^{\rm bulge}_{\rm gas}(\rm M_{\odot})$ & $1.2 \cdot10^{10}$  & $1.0 \cdot10^{9}$ &  $2.2 \cdot10^{10}$ \\
  \hline
$M^{\rm disc}_{\rm gas}(\rm M_{\odot})$ & $2.8 \cdot10^{10}$  & $1.5 \cdot10^{10}$ & $2.9 \cdot10^{10}$  \\
  \hline
$M^{\rm disc+bulge}_{\rm gas}(\rm M_{\odot})$ & $4.0 \cdot10^{10}$  & $1.6 \cdot10^{10}$ & $5.1 \cdot10^{10}$ \\
  \hline
  \hline
$M^{\rm bulge}_{\rm DM}(\rm M_{\odot})$ & $5.9 \cdot10^{9}$  & $1.1 \cdot10^{9}$ & $1.2 \cdot10^{10}$  \\
  \hline
$M^{\rm disc} _{\rm DM}(\rm M_{\odot})$ & $5.5 \cdot10^{10}$ & $2.9 \cdot10^{10}$ & $1.1 \cdot10^{11}$  \\
  \hline
$M^{\rm disc+bulge}_{\rm DM}(\rm M_{\odot})$ &$6.1 \cdot10^{10}$  & $3.0 \cdot10^{10}$ & $1.2 \cdot10^{11}$  \\
  \hline
  \hline
$\lambda_{\rm DM}({\rm spin})$ & $0.048\pm 0.002$ & $0.051\pm 0.002$ & $0.107\pm 0.012$  \\
  \hline
\end{tabular}
\label{tab:feature}
\end{table}

We assume a flat $\Lambda$CDM cosmology with total matter
density $\Omega_{m}=0.27$, baryon density
$\Omega_b=0.045$, dark energy density $\Omega_{\Lambda}=0.73$,
amplitude of the matter power spectrum $\sigma_8=0.8$ and 
Hubble constant $H_0=70\, \rm km \cdot s^{-1} \cdot \rm Mpc^{-1}$
consistent with the WMAP 7-year data \citep{komatsuetal11}.  The
box size of our simulations is $L_{\rm box}=100\,  h^{-1}\cdot \rm Mpc$ 
for both simulations.  We generate high-resolution initial conditions  within a small
Lagrangian region enclosing all dark matter (DM) particles that end up within  $2\, r_{\rm
vir}$ of the halo at $z=6$ . The DM particles have a mass  of
$M_{\rm res}=9\cdot\,10^{5}\, h^{-1}\cdot \rm M_{\odot}$  and  $7\cdot\,10^{6}\,
h^{-1}\cdot \rm M_{\odot}$ in the SH and LH simulation,
respectively.  That corresponds to an equivalent $4096^3$ cartesian
grid for the SH (and SHhr) simulation, and a $2048^3$ grid for the LH simulation. 

These simulations were  run with the AMR code {\sc ramses}
\citep{teyssier02}. The evolution of the gas was followed using a
second-order unsplit Godunov scheme for the Euler equations. The
HLLC Riemann solver with a first-order MinMod Total Variation Diminishing
scheme to reconstruct the interpolated variables from their
cell-centered values was used to compute fluxes at cell interfaces. Collisionless particles (dark matter and star
particles) were evolved using a particle-mesh solver with a
Cloud-In-Cell (CIC) interpolation.

The initial mesh is refined with up to 9 levels of refinement. The
maximum spatial resolution is chosen to be constant at a value  $\Delta x=95\, h^{-1}\cdot \rm
pc$  in physical coordinates.  We also  performed  a higher resolution simulation with 12 levels
of refinement, corresponding to $\Delta x=12\, h^{-1}\cdot \rm pc$ for the
lower mass halo (SHhr). Grid  cells are refined following a
quasi-Lagrangian criterion if more than 8 DM particles lie
in a cell, or if the baryon mass exceeds 8 times  the initial dark
matter mass resolution.

Gas cooling is implemented assuming 
a fixed metallicity of $10^{-3} Z_{\odot}$ with a Solar composition for the gas \citep{sutherland&dopita93}.
Heating from the UV background follows \cite{haardt&madau96} with
reionisation occurring at $z_{\rm reion}=8.5$.
The cooling module takes into account metal line cooling, allowing gas temperatures 
to reach a 100K temperature floor. 

Star formation is assumed to occur in regions where the gas
density reaches $n_{\rm H} > n_0= 1 \, \rm H \cdot cm^{-3}$ for the SH and LH runs
($n_0=50\, \rm H \cdot cm^{-3}$ for the SHhr run) and a random
Poisson process is used to spawn star cluster particles (see
\citealp{rasera&teyssier06} and \citealp{dubois&teyssier08winds} for
more details). The star formation rate is assumed to scale according to the 
 Schmidt-Kennicutt law $\dot \rho_*= \epsilon_* {\rho / t_{\rm ff}}\, ,$ where $\dot \rho_*$ is the star
formation rate density, $\epsilon_*$ the star formation efficiency,
and $t_{\rm ff}$ the local free-fall time of the gas. We set the efficiency of
star formation to be $\epsilon_*=0.01$ in good agreement with 
observed star formation surface densities in galaxies \citep{kennicutt98}, and 
observations of local giant
molecular clouds \citep{krumholz&tan07}.  Star cluster particles are
given a mass of $m_*=\rho_0\Delta x^3$, with $\rho_0=n_0 m_{\rm p}/X_{\rm H}$, $m_{\rm p}$ is the proton mass, and $X_{\rm H}=0.76$ is the Hydrogen fraction. The stellar mass resolution reaches $5.5 \cdot 10^4\, h^{-1}\cdot \rm M_{\odot}$ for the low resolution simulations (SH and LH) and $5.5 \cdot10^3\,
h^{-1}\cdot \rm M_{\odot}$ for the high resolution simulation (SHhr). We do not account for the effect of
feedback due to  supernovae (SNe) or AGN  as we want to  focus here on the
infall  of material from the cosmic web. We leave the effect of
feedback for further work. 

The top panels of fig.~\ref{fig:visual} show the
setting of the gas around the two halos at $z=6$.
Most of  the gas  flows in
cold  along pronounced filamentary streams, and an accretion shock
develops at the virial radius, increasing the amount of hot gas being
accreted with time \citep{birnboim&dekel03,keresetal05,ocvirketal08,brooksetal09,dekeletal09}.  
The more massive halo in the LH simulation has
a larger shock-heated region than the lower mass halo in the SH simulation, but nevertheless  shows
a cold filamentary accretion component penetrating the virial sphere as expected for this mass and redshift. Note that the stability of the shock is determined by the shape of the cooling curve and, thus, by the level of enrichment of the Intergalactic Medium (IGM)~\citep{birnboim&dekel03}. Here, as we do not include any SN feedback, the value of the metal enrichment of the IGM is somewhat lower than what is expected, and the shock is more stable. However, SNe inject thermal energy to the surrounding gas and it is not clear how this would modify the stability of the shock.
The bottom panels of fig.~\ref{fig:visual} are zooms of the
high-resolution simulation  SHhr  which show a
gravitationally central clumpy disc of gas (bottom left), and stars (bottom right). 
Note, however that as we will
discuss later  a significant fraction of the gas   streams directly
into the compact bulge at the centre without ever settling into a
disc. 

The black curves in fig.~\ref{fig:vrot_vcirc} show the contribution of  gas, stars and DM to  the circular velocity curves $v_{\rm circ} = \sqrt {GM/r}$  of the two halos while the red curves show the measured rotational velocities, and the blue curves show the radial velocity dispersion. 
In both halos the baryons flowing to the centre become self-gravitating at a radius of about 3 kpc. 
The compact density distribution of the stars thereby leads to a significant inward rise of the circular velocity, with the rotational velocity (red dot-dashed lines) of the gas closely matching the circular velocity of the total mass distribution in the core of the halo $r<0.1\, r_{\rm vir}$. 
The higher resolution SHhr simulation nicely resolves the peak of the rotation curve which reaches 900 km/s at a radius of 70 pc, that corresponds to a very compact bulge.
Note that the upwards jump in the circular velocity curve for the LH simulation is due to the  presence of a satellite galaxy  at a distance of 15 kpc from the centre of the main galaxy. 

We apply a bulge/disc decomposition by fitting two exponentially decreasing profiles on the the stellar surface density (see Appendix A of~\citealp{duboisetal12}). 
This decomposition provides us with a stellar bulge (disc) mass $M_{\rm star}^{\rm bulge}$ (resp. $M_{\rm star}^{\rm disc}$). 
To infer the mass of gas (DM) into the bulge $M_{\rm gas}^{\rm bulge}$ (resp. $M_{\rm DM}^{\rm bulge}$), we measure the radius that encloses $M_{\rm star}^{\rm bulge}$ and compute the gas (resp. DM) mass content within this radius. 
We apply the same procedure for the total (disc$+$bulge) mass of gas (DM), and get the disc component by differentiating $M_{\rm gas}^{\rm disc+bulge}$ with $M_{\rm gas}^{\rm bulge}$ (resp. $M_{\rm DM}^{\rm disc+bulge}$ with $M_{\rm DM}^{\rm bulge}$).
The central bulge contains 14, 24 and 41 \% of the total baryonic mass  for the LH, SHhr and SH
simulation, respectively (see table~\ref{tab:feature}). The gas fraction within it is 30 \% for the
two lower resolution and 5\% for the higher resolution simulations. 
The decrease of the gas fraction with increasing resolution is expected due to denser clumps with lower free-fall time that form stars faster.
Note that the stellar bulge rotates significantly faster than the DM but the  rotational velocities are typically still a factor 2-4 lower than the circular velocity. 
The stellar bulge, while somewhat flattened is thus far from rotational support, and is supported by the radial velocity dispersion $\sigma_{*}$. 
At $0.1<r<2$kpc the compact central bulge is surrounded by a massive gravitationally unstable and rotationally  supported gas disc.
The stellar disc of the SHhr galaxy at $z=6$ is supported by its rotation and its velocity dispersion, which explains its large disc scale height ($\sim 0.5$ kpc) compared to the gas disc component ($\sim 0.1$ kpc), and its old average formation time (243 Myr) compared to its dynamical time (3 Myr).
The DM is essentially non-rotating and get its support against gravitational collapse  from the radial velocity dispersion $\sigma_{\rm DM}$ of DM particles.

\begin{figure*}
   \includegraphics[width=1.0\columnwidth]{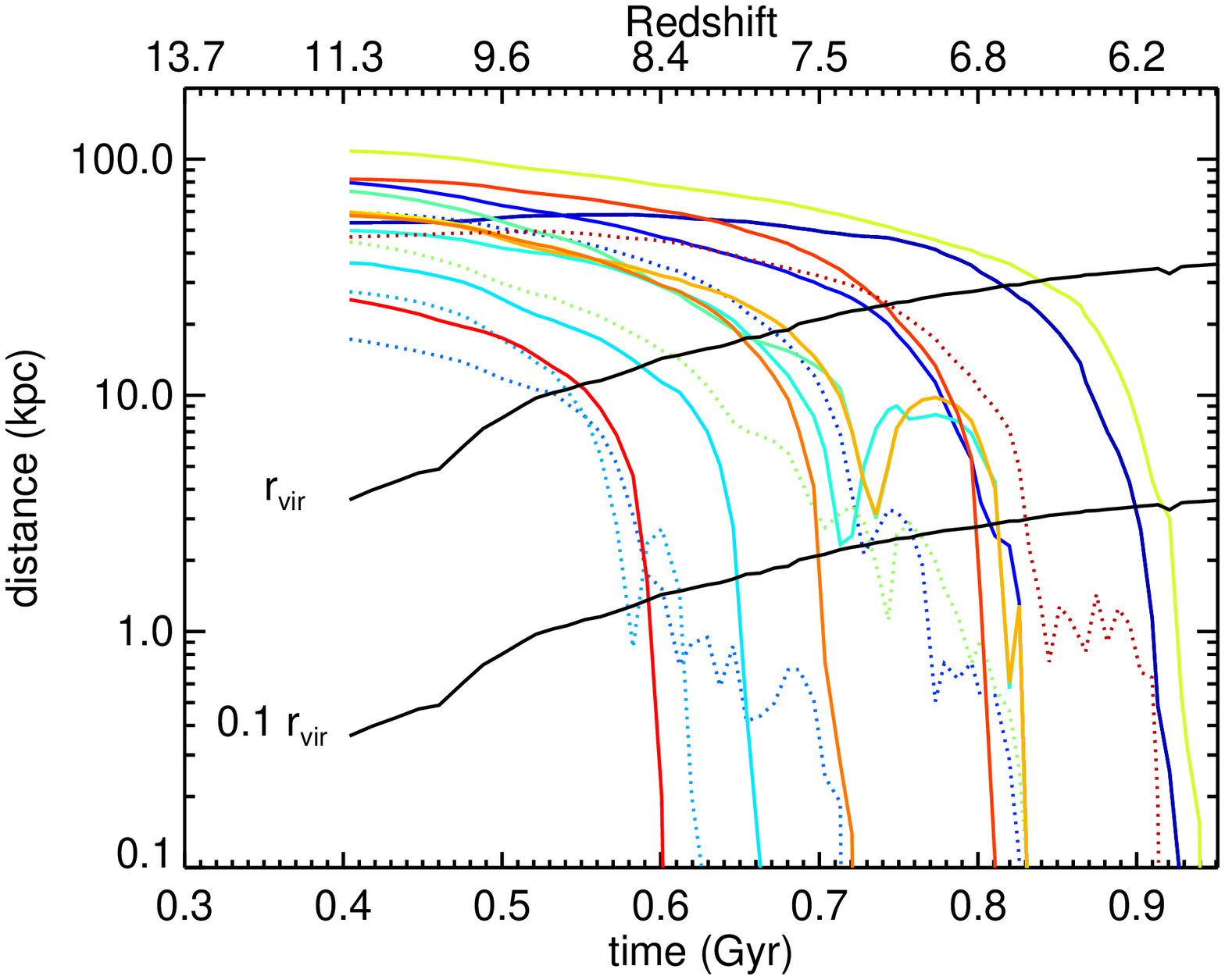}
   \includegraphics[width=1.0\columnwidth]{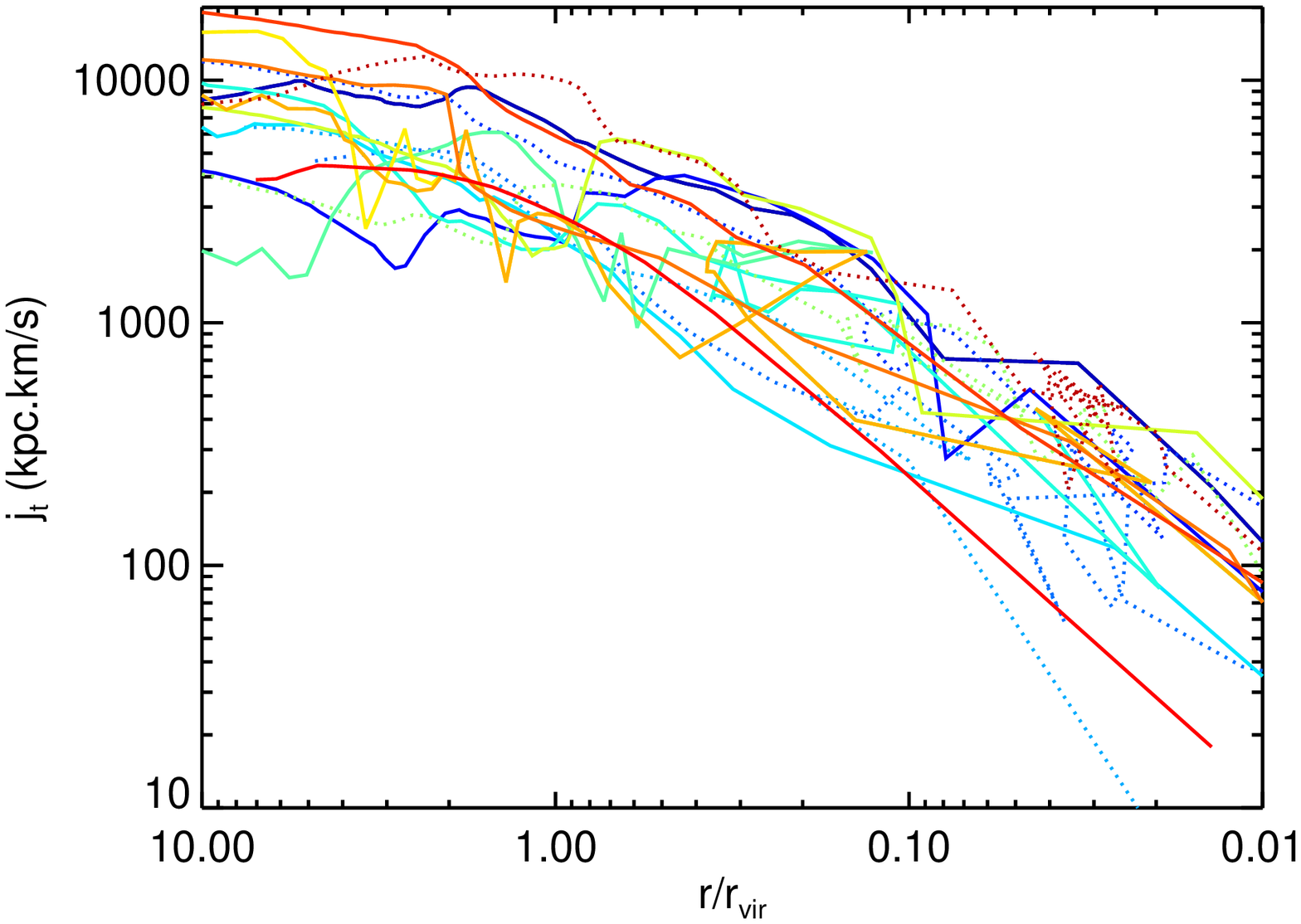}
      \caption{{\it Left panel:}  Distance from the galaxy centre of 15 tracer particles
randomly selected from the bulge tracer particles identified at $z=6$ in the SHhr halo. 
Solid curves are for particles tracing matter directly accreted from the circum-galactic medium (CGM) into the bulge, whereas dotted
curves trace matter accreted through the central disc. 
Solid black curves show  $r_{\rm vir}$ and $0.1\, r_{\rm vir}$. Note that the tracer particles always rapidly sink as they enter $r_{\rm vir}$, but sometimes linger in the CGM around $0.1\, r_{\rm vir}$.
{\it Right panel:} specific angular momentum of the same tracer particles as in the left panel 
relative to the center of the bulge  as a function of the distance to the center of the galaxy.
}
\label{fig:disc_cgm_selection}
\end{figure*}

Let us close this section with a brief discussion how we will
investigate the geometry of the cold gas inflow into the halos
down to $z=6$. We make use of massless \emph{tracer} particles
\citep[e.g.][]{mitchelletal09}   to circumvent  the
difficulties  that  grid-based methods pose  for   following  the
Lagrangian path of the gas through the cosmic web into collapsed
structures. In contrast to standard particles (DM or stellar
particles), tracer particles are advected by using the updated
velocity of the gas cells with the Riemann solver. Each tracer
particle has its own velocity computed from a CIC
interpolation. Tracer particles are uniformly distributed on the
initial mesh of the simulation. We checked that the tracer particles rotational velocity profile closely matches its gas counterpart (see Appendix~\ref{App:tracer}).
  Figure \ref{fig:visual-tracer}
displays the particles of the SH simulation  which are identified to
belong to the bulge of the central galaxy  at $z=6$ as they are distributed at
$z=9$. The filamentary origin of the bulge particles is very
clear, confirming that a large fraction of the gas does indeed
stream to the centre of the halos along the prominent filaments
making up the cosmic web.

\section{Mass accretion and angular momentum transport in the halo}
\label{sec:mass}

\subsection{Tracking the evolution of angular momentum}

We are mainly interested here in the gas settling in the central region
of the massive halos.  This gas thus necessarily has low angular
momentum.  We therefore first  compare the angular momentum of the
gas falling to the inner bulge region of the galaxy (as defined by the scale length 
of the exponential fits performed for the bulge/disc composition: 200 pc for the SH
and LH runs, and 40 pc for the SHhr run, see fig.~\ref{fig:visual}) to
that of  the gas settling into the central disc surrounding the
bulge. 

In order to do this, we use the tracer particles to follow back in time the angular
momentum of gas that ends up at a given redshift in either the
central bulge or  central disc  and we calculate its specific angular
momentum  relative to the centre of mass, $ \mathbf{\overline r}$, of
the same (large) ensemble  of  tracer particles  
($\sim 2.3\, 10^5$ tracer particles located in the bulge of the SH galaxy at
$z=6$).  The green, red and black solid curves   in
fig.~\ref{fig:momemtumvsz}  show the specific angular momentum
evolution of  gas identified in the bulge  at $z =$  9, 7 and 6,
respectively. Note that we have truncated the curves when the gas has
reached  the resolution limit of the simulation.  The dashed curve is
the evolution of the angular momentum of  the gas that has settled
into the surrounding disc with $0.2 \, {\rm kpc} < r < 2
\, \rm {kpc}$ at $z=6$.    We also show the specific angular momentum of
the DM as well as the specific angular momentum of the tracer particles within $r_{\rm vir}$ at $z=6$  in
the two massive halos.  In each case we have calculated the angular
momentum as  $j\equiv | \sum_i (\mathbf{r}_i-  \mathbf{\overline
r})\times (\mathbf{v}_i-  \mathbf{\overline v})|$. The mean angular
momentum of the DM  and the gas  in the halo grow in
reasonable agreement with  tidal torque theory but, as discussed in
detail by  \cite{kimmetal11}  and \cite{pichonetal11} in simulations
with gas cooling,  the angular momentum of the gas
in the outer parts  of the halo ($0.1\, r_{\rm vir}<r<r_{\rm vir}$) 
is higher than that  of the DM by a factor of 2-4, 
decreasing with increasing mass or redshift and thus rareness of the halo.  In our simulation we find a
factor 2 difference similar to the more massive halos of  \cite{kimmetal11}  and \cite{pichonetal11}. 
As shown by the  dotted lines in   fig.~\ref{fig:momemtumvsz},  the average specific 
angular momentum  of the baryons (gas + stars) and DM within $r<r_{\rm vir}$ in the two halos is  comparable.

The specific angular momentum of tracer particles ending up in the 
disc of the central galaxy also appears to grow in
at late times. The specific angular momentum oscillates and globally decreases rather than increases
due to gravitational perturbations (clump migration, mergers, etc.) which lead to inward mass
and outward angular momentum transfer in part to
the outer parts of the disk presumably, but most likely to the stellar and
diffuse gaseous component of the halo.  However, we are mainly interested  here
in the gas with the lowest angular momentum which forms the bulge.
Initially the specific angular momentum of the gas ending up
in this central bulge is lower than that settling into the disc, up to 
 a factor two in the case of the more
massive halo. At later times, when most of the mass actually
assembles into the bulge, its evolution decouples from that of the
central disc and its gas loses
another factor $5-10$  in angular momentum. 

 In the following we use tracer
particles to investigate in detail the trajectories  of the flow of
gas towards the bulge and discuss how it achieves sufficient 
angular momentum loss to settle into as compact a  bulge as observed in our
simulations. 
\begin{figure*}
 \includegraphics[width=0.6\columnwidth,angle=90]{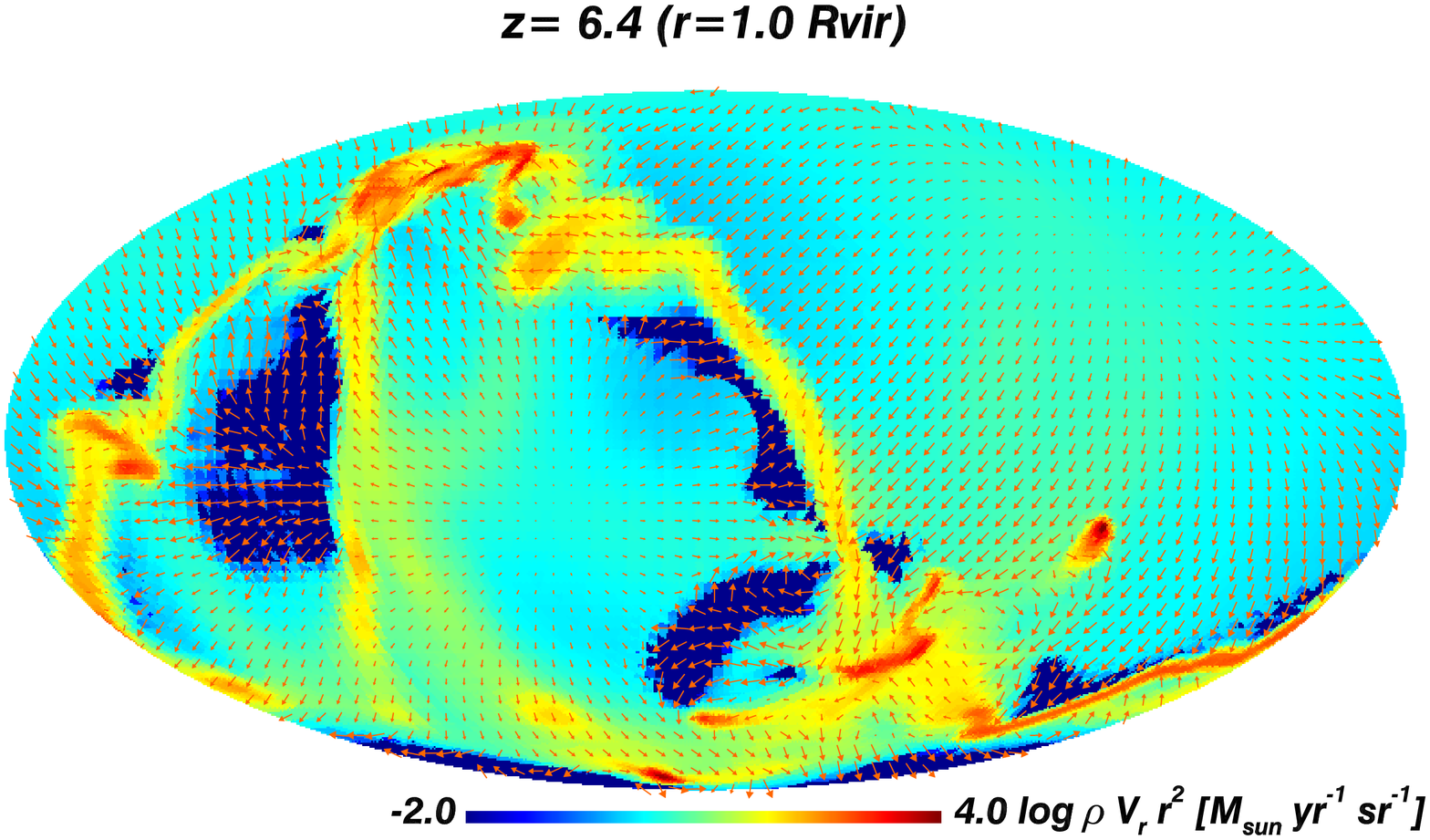}
  \includegraphics[width=0.6\columnwidth,angle=90]{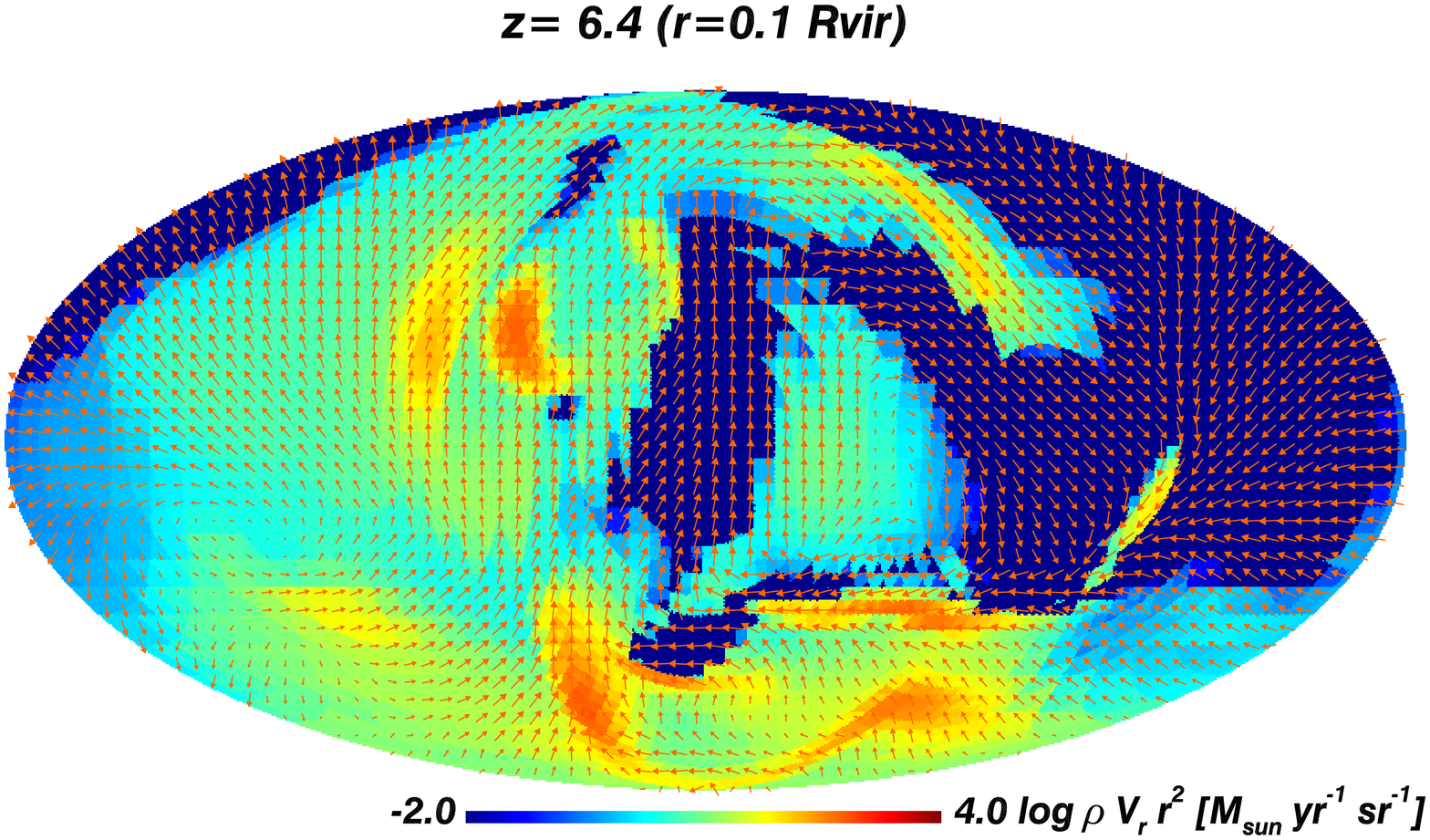}
  \includegraphics[width=0.6\columnwidth,angle=90]{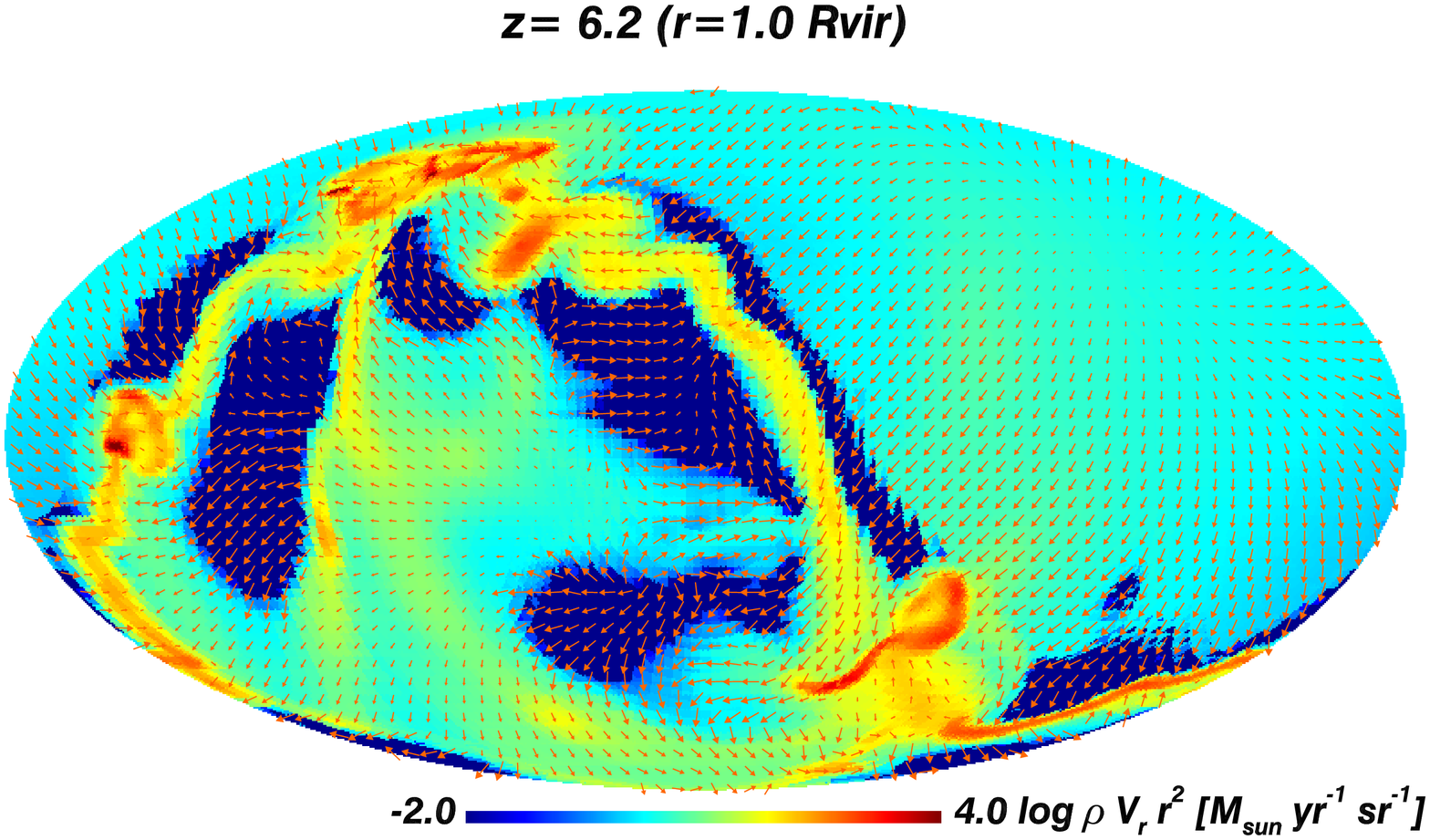}
  \includegraphics[width=0.6\columnwidth,angle=90]{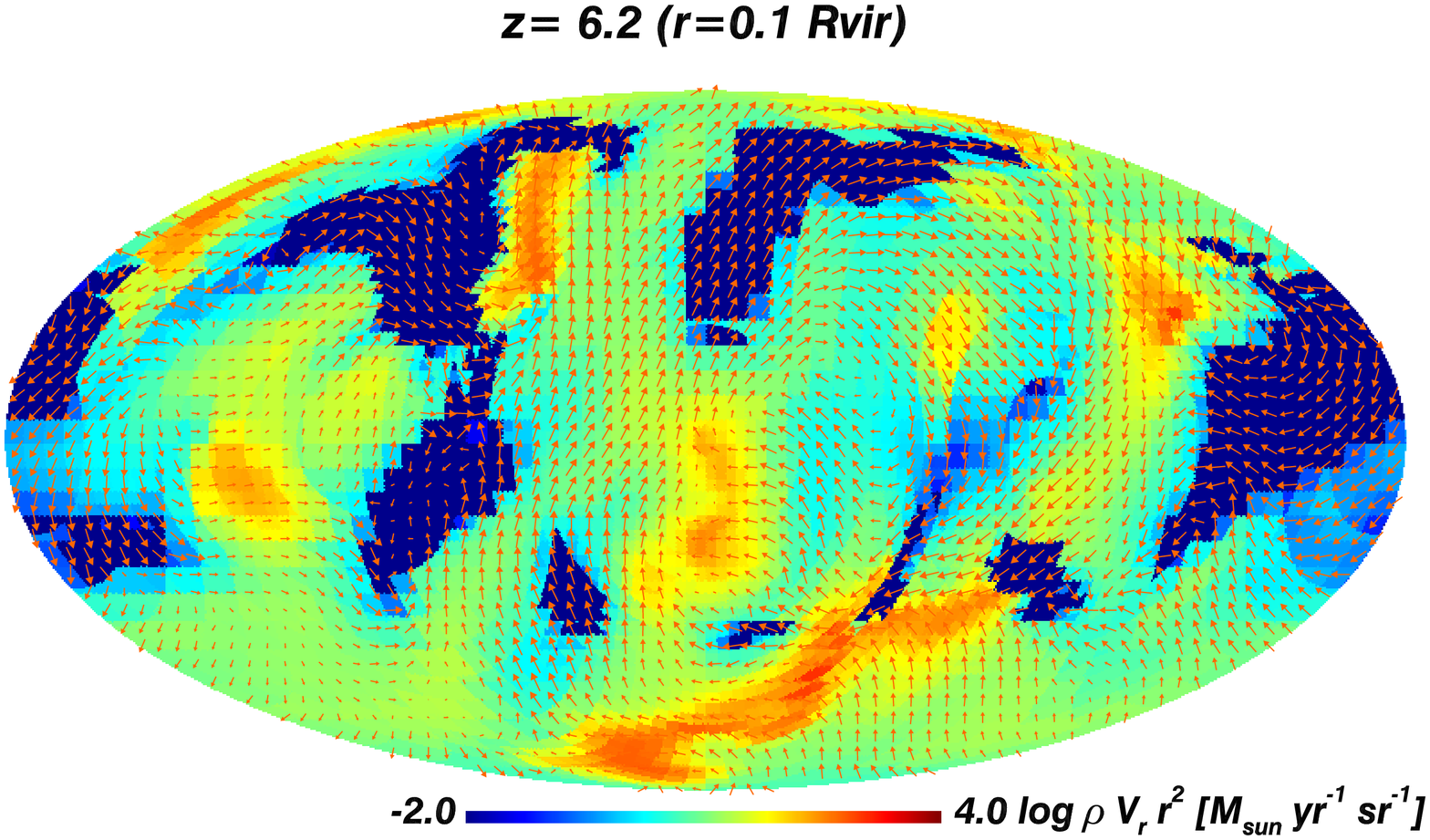}
  \includegraphics[width=0.6\columnwidth,angle=90]{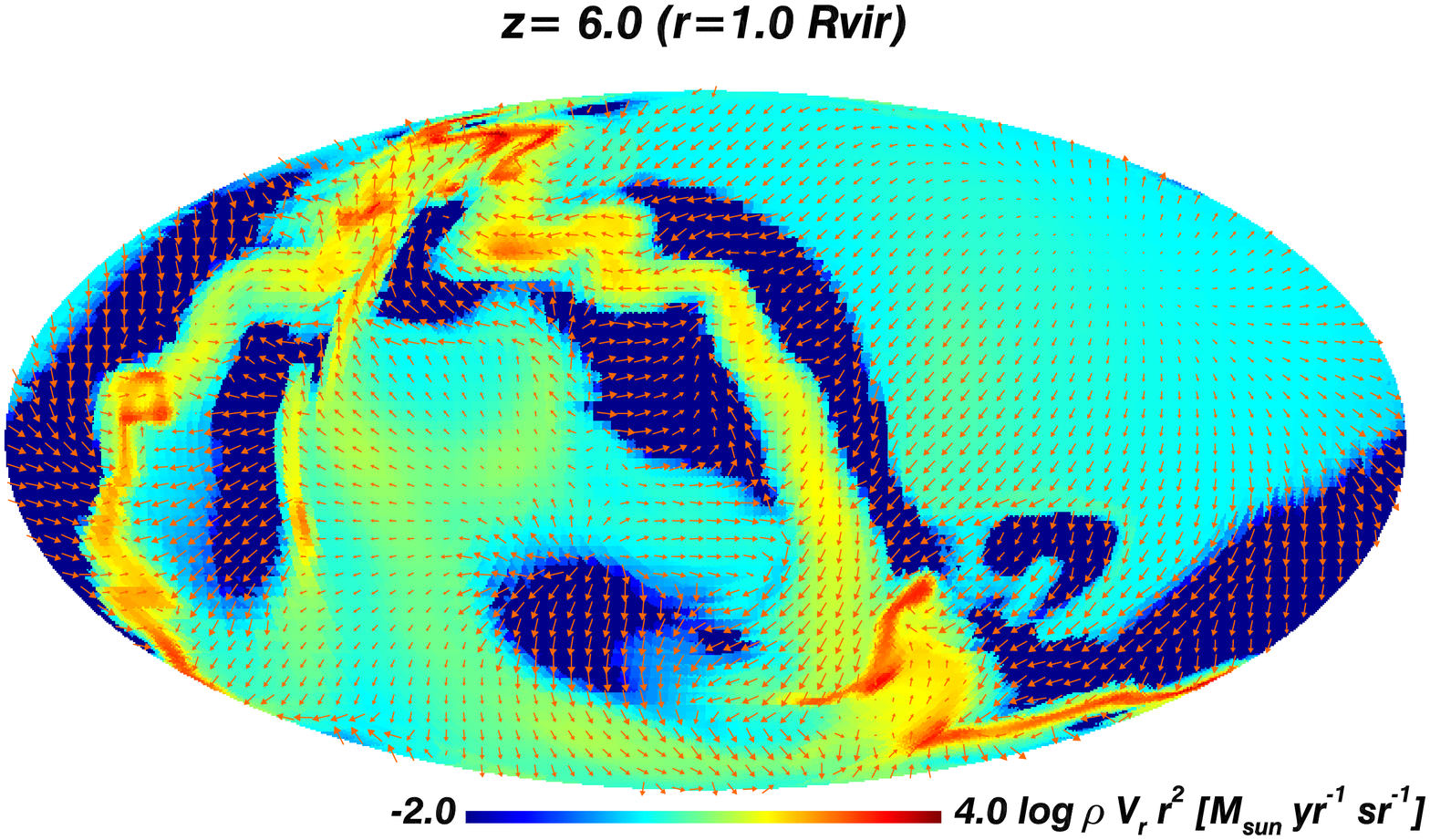}
  \includegraphics[width=0.6\columnwidth,angle=90]{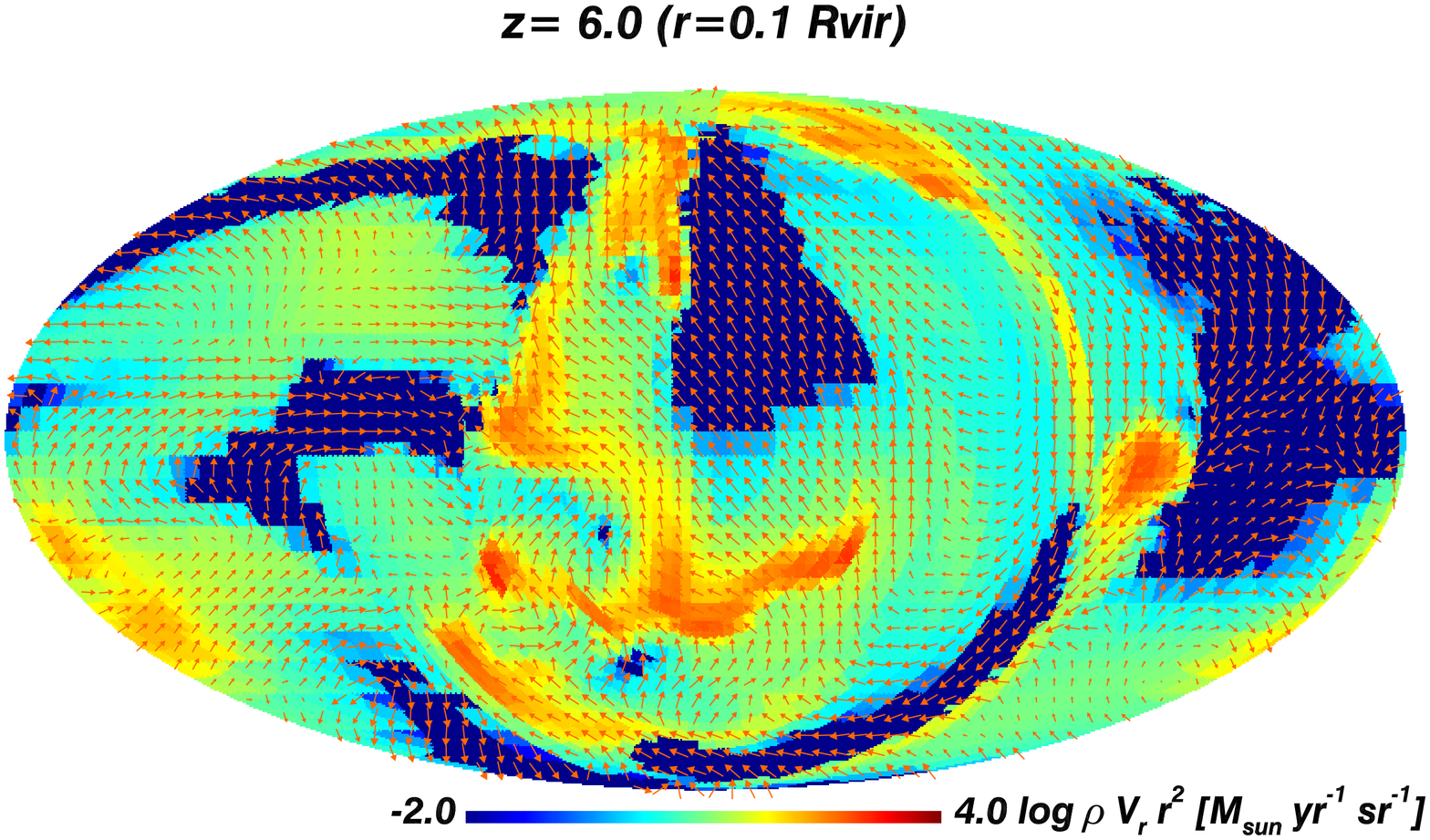}
       \caption{  {\sl  Left panels: } Angular distribution of the mass flux  of gas passing
         through the virial radius $r_{\rm vir}$ together with the 
          tangential velocity field at  $z=6.4, 6.2$ and $6$ ({\sl
            from top to bottom}). Note the consistency of the characteristic patterns of sheets
           feeding the halo, with the largest inflow rates located at the
           position of the filaments, i.e. at the intersections of the sheets. Note also the coherent 
           tangential velocity field on large angular scales with matter
           flowing away from the voids into the walls and along the 
           filaments.  {\sl   Right
           panels:} Same  at $0.1 \, r_{\rm vir}$. The angular distribution at the
           smaller radius changes rapidly with time, reflecting the complex non-linear dynamics
            within the core of the halo. These maps are constructed from the high resolution SHhr simulation.}  
\label{fig:healpix_map}
\end{figure*}

\begin{figure*}
  \includegraphics[width=0.66\columnwidth]{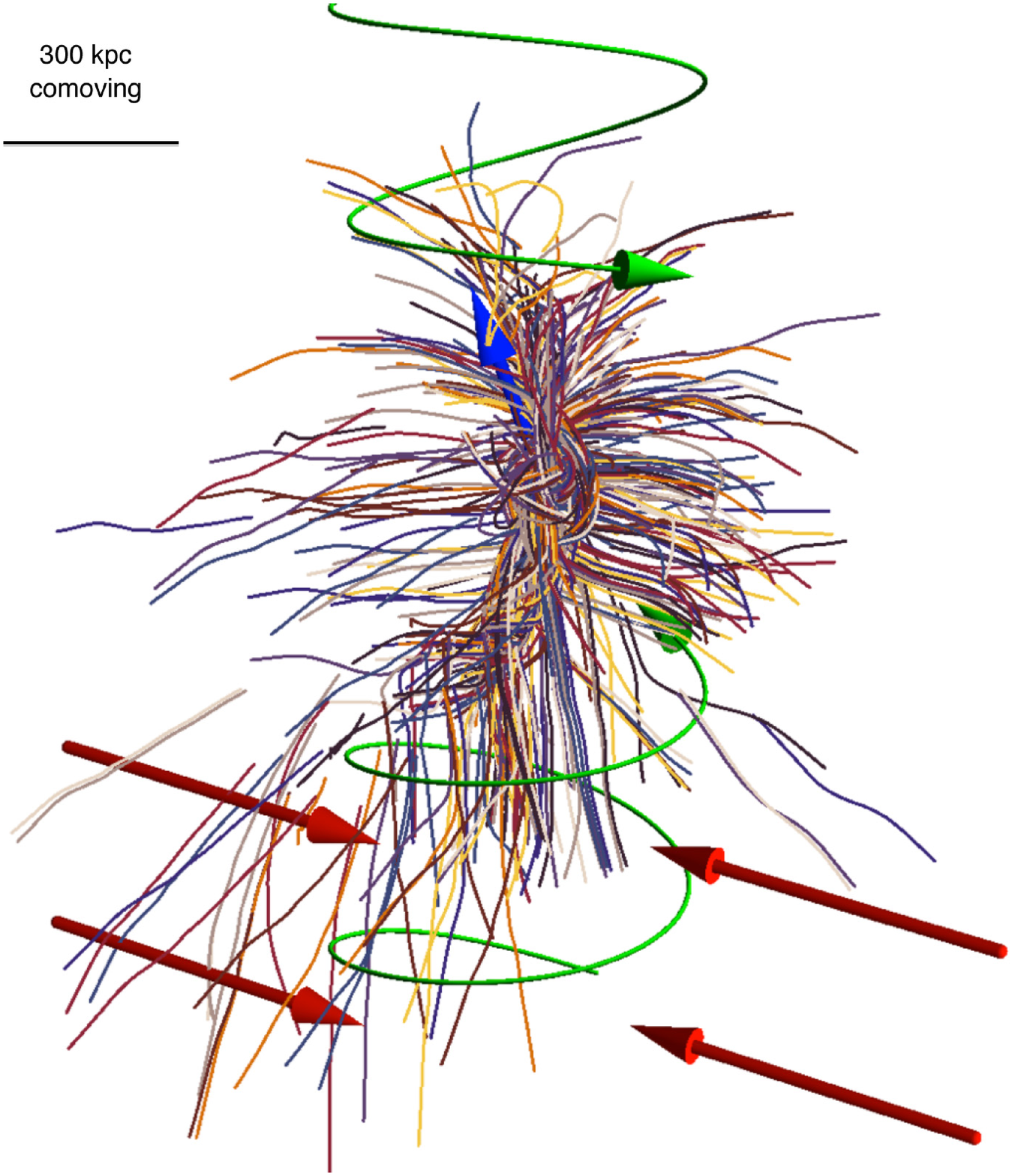}
  \includegraphics[width=0.66\columnwidth]{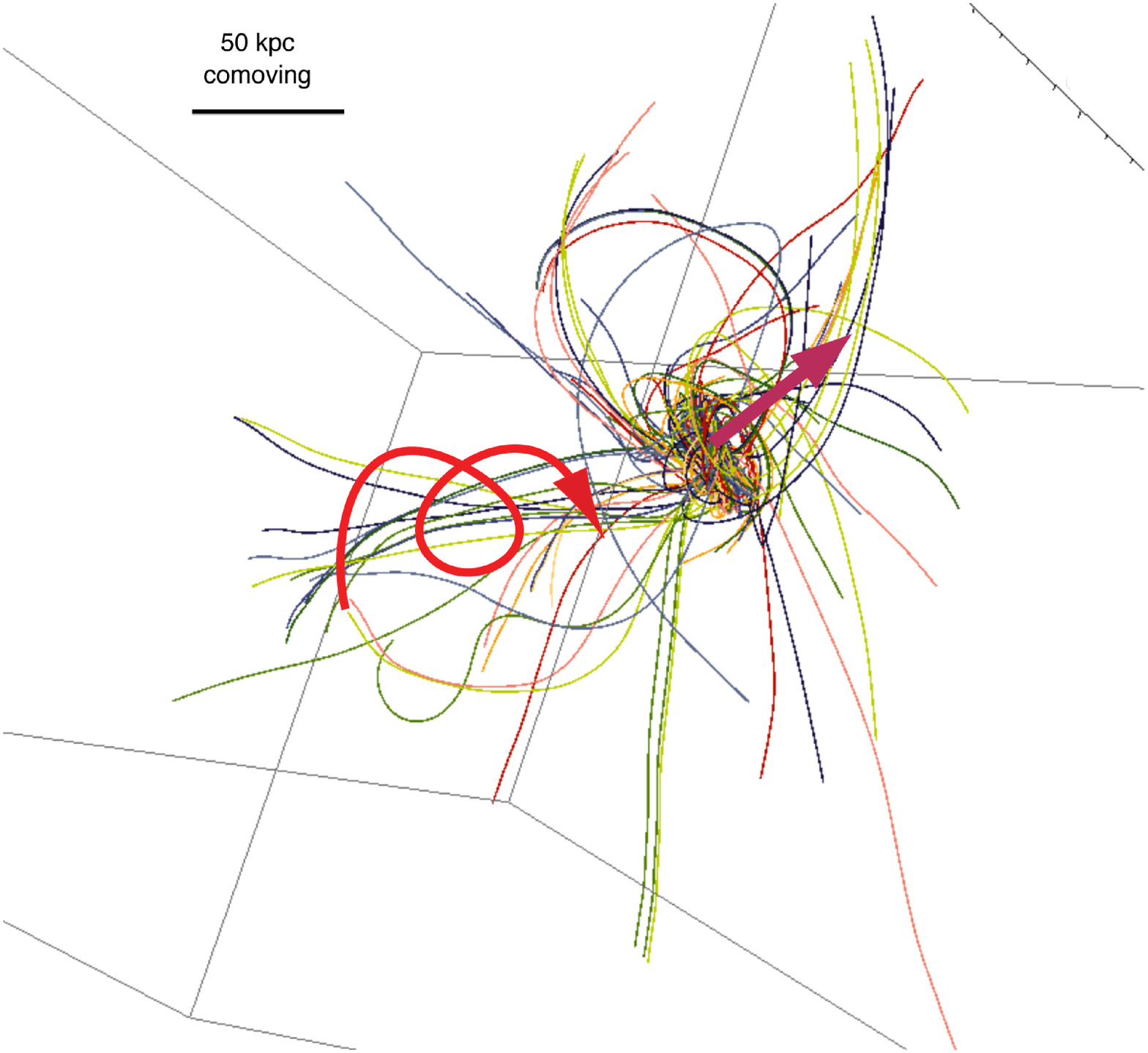}
  \includegraphics[width=0.66\columnwidth]{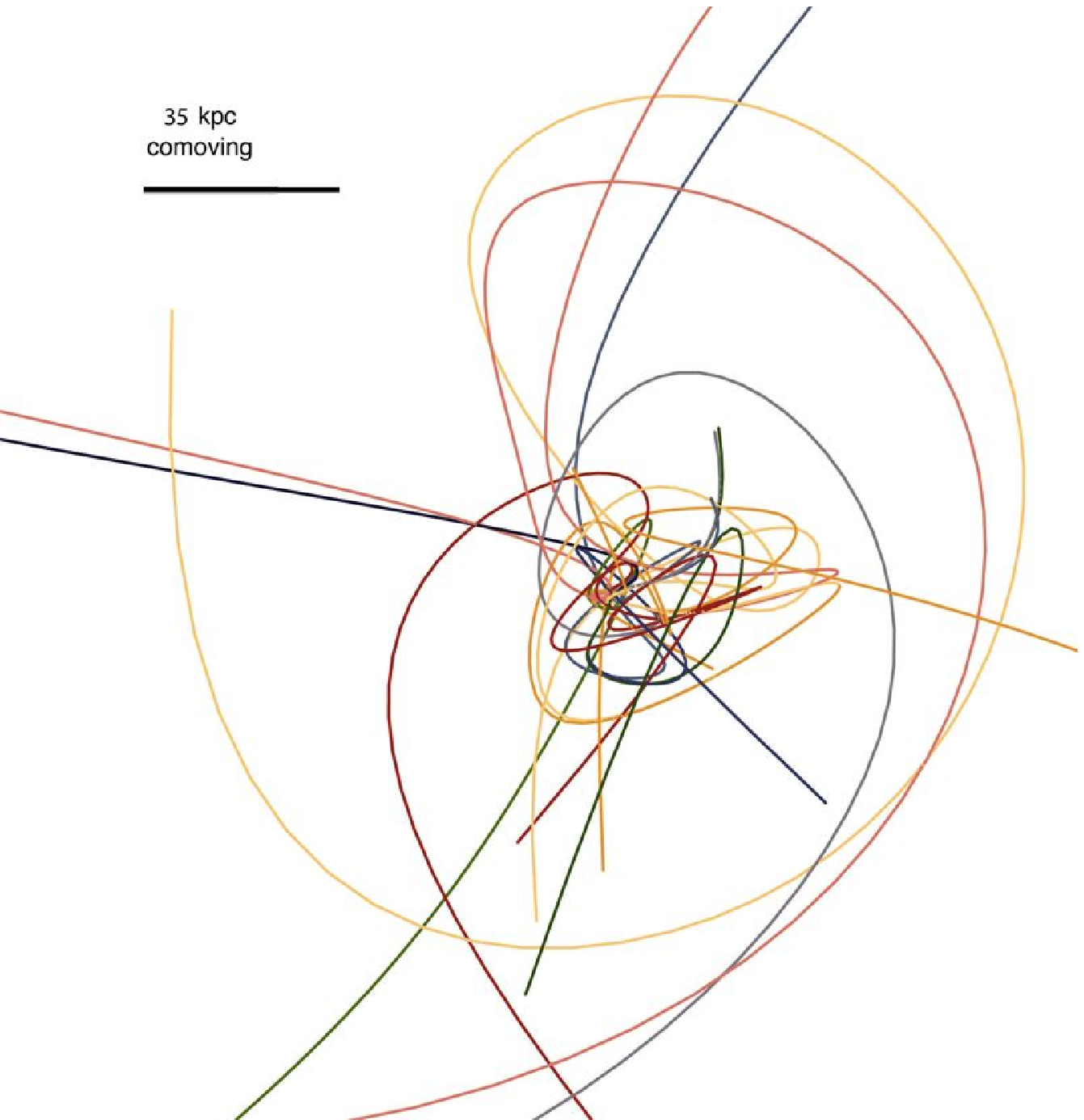}
      \caption{The trajectories of bulge tracer particles  in the outer
        region ({\sl left panel})  on intermediate scales ({\sl middle panel}) and 
     in the inner region of the halo ({\sl right panel}). On super halo scales, the gas
     flows out of the  voids into the the walls/sheets  (red arrows) and winds
     around the  filaments ({\sl green arrows}, see also
     fig.~\ref{fig:healpix_map}). The 
     gas and proto-galactic clumps flow along the filaments
     and  merge with the most massive clump  which will later become the compact central bulge
     ({\sl middle panel}).
     A large fraction of the gas is funneled directly to the compact bulge at
     the centre,  while the rest settles briefly into the surrounding
     disc. The disc   gets  continuously disturbed by merging  clumps 
     and is thus unstable and transient. As a result, the gas in the disc  rapidly loses angular
     momentum and contributes to  the growth of the  gas rich bulge 
    ({\sl right panel}; the horizontal bar corresponds to the Virial radius of SHhr at $z=6$).
    }
\label{fig:tracer-orbit}
\end{figure*}

\subsection{Building up   bulge from  baryons in the cosmic web}

Let us now take a closer look at  fig.~\ref{fig:visual-tracer} and
interpret it taking advantage of the fact that we can follow the
trajectories  of the gas flow using tracer particles.  The  ensemble of these
particles ending up at the centre of the galaxy is initially distributed in a
sheet-like structure  within which is embedded a dominant  filament. The tracer
particles flow  from this sheet into filaments where they form
"protogalaxies". The  gas  in the sheet typically has a non zero impact
parameter relative to the filament  axis and protogalaxies thus acquire
spin parallel to the filament in which they form \citep[see also][]{codisetal12}.  The gas then
streams along  these filaments into the dominant halo forming at the
centre of the zoom region. These cold filamentary streams  
directly  feed the central disc and bulge discussed in the  last
section. The protogalaxies falling in along these streams lead to
minor mergers and are dragged in by dynamical friction and
torquing. The  (orbital and spin)  angular momentum  of the inflowing gas  and that of the
merging protogalaxies originating from different directions  thereby add up incoherently (vector cancellation).
Part of the inflowing matter directly feeds the compact bulge
and another part settles temporarily  into the surrounding disc. This is
nicely demonstrated in  the left panel of fig.~\ref{fig:disc_cgm_selection} which shows
the distance to the galaxy centre of 15 randomly chosen tracer
particles ending up in the bulge at $z=6$ as a function of time.   The particles enter the
halo  and once they pass the virial radius, they quickly fall to the centre.
A sizable fraction of them hangs around $0.1\, 
r_{\rm vir}$ for a while and settles into this  gravitationally unstable central
disc,  but more than half
fall straight into the central bulge. 
We also see that the gas that settles into the bulge later in time comes from
regions on the outskirts of the initial patch.
The right panel of fig.~\ref{fig:disc_cgm_selection} also shows 
that outside the virial radius of the halo, the particles that originated from further away generally tend
to have a larger angular moment. The evolution of the angular momentum of the tracers
has a significant stochastic component, but once they cross $r_{\rm vir}$, 
these tracers of the bulge gas are seen to lose the angular momentum 
while sinking in.

Figure~\ref{fig:healpix_map} shows the angular distribution of the
mass influx  at the virial radius on the left-hand side and deep
within the halo at a tenth of the virial radius on the right hand 
side in the form of  Mollweide maps (constructed using  {\sc \small
 HEALPix} \citealp{hivon05}) at three redshifts
$z=(6.4,6.2,6.0)$. The maps  also show the tangential velocity field 
in the form of arrows (the longer the arrow, the larger the tangential velocity). At the  virial radius the angular pattern is
very stable reflecting the inflow from the sheets and filaments
constituting the large scale cosmic web. The tangential velocity field 
shows the expected drift of matter from voids into sheets/walls (the 2D
bridges, e.g. north-west to south-east in left panels) and  filaments 
\citep[the 2D peaks on these maps][]{pichonetal11, danovichetal12}. In contrast, the flow at $0.1 \, r_{\rm
 vir}$ is much more stochastic both in time and direction, and is a
signature of a rapid redistribution/segregation  of angular momentum 
within the collapsed halo~\citep{kimmetal11}.

As a next step, let us now take a look at the geometry of the trajectories 
of the tracer particles as they stream together with the gas from the
cosmic web into the bulge. In fig.~\ref{fig:tracer-orbit} we show 
a  selection of such trajectories starting at $z=9$ for the 
SHhr simulation  on three different scales as indicated on the plot.
In the three panels a small subset
of tracer particles are chosen randomly and are followed for a
fraction of the remaining simulation time. Time increases from left to right panel.  
On large scales (at early times), we note that the  flow is indeed
dominated  by the winding of gas streaming from the main wall visible in
fig.~\ref{fig:visual-tracer} around the main filament.  The tracer
particles start perpendicular to the filament within the walls.  As
they  reach the filament, they  take a (sharp) turn loosing their transverse motion and
flow along the filament (left panel of fig.~\ref{fig:tracer-orbit}). 
 This  complex structure  converges
into  a  quite narrow and  elongated ``plait'' on either side of the
forming disc.  The disc will advect
further infalling gas  at its periphery  preferentially along its spin axis
(middle panel of fig.~\ref{fig:tracer-orbit}).  The gas falling towards  the disc from 
the Circumgalactic Medium (CGM) also shocks as it 
 approaches the  disc and takes (another) sharp turn aligning
itself with the (transient)  spinning disc.  This flow drags along satellites (collapsed
structures/proto-galactic clumps  that formed outside  the halo in the cosmic web) and cold filamentary non-star forming gas (right panel of fig.~\ref{fig:tracer-orbit}). 

\subsection{The radial evolution of fluxes}

Let us define different phases for the gas:  the star forming gas with $\rho \ge \rho_0$, the hot diffuse gas phase with temperature $T\ge10^5$ K and $\rho<\rho_0$, the cold filamentary phase with $T<10^5$ K and $100\, \bar \rho_{\rm b}\le\rho<\rho_0$ ($\bar \rho_{\rm b}$ is the average baryon density of the Universe), and the cold diffuse phase with  $T<10^5$ K and $\rho<100\, \bar \rho_{\rm b}$.
A careful inspection of the mass flux at $0.1 \, r_{\rm vir}$ and $r_{\rm vir}$ shows that the infall in the SHhr halo is largely dominated by the cold filamentary (non-star forming) gas with high accretion rate $\dot M_{\rm acc}=100-200 \, \rm M_\odot$ (left and right panels of fig.~\ref{fig:fluxSHhr}), high enough to bring sufficient amounts of gas that drives a recurrent Toomre unstable disc (bottom left panel of fig.~\ref{fig:visual}) with typical Toomre parameter of the order of $Q\sim 1$ in the central galaxy of the SHhr simulation \citep[see also][]{dekeletal09clumps, ceverinoetal10}.
Multiple clumps are observed in the disc and very few are accreted through mergers ({\sl ex situ}). 
When measured at $r_{\rm vir}$ for the SHhr halo at $z=6$, 10 \% of the total accreted mass of gas is in the form of compact star forming clumps, 55 \% is accreted through cold streams, and the rest 35 \% is a diffuse accretion.
This proves that clumps form within the galaxy ({\sl in situ}) via strong disc instabilities (see also fig.~\ref{fig:clump_migration} below).
This process occurs generically, but for rare peaks, the larger number of connected filaments, and the importance of minor merger accelerates the radial migration. 
This rapid inflow of fresh gas is the result of comparable levels of accretion at $r_{\rm vir}$, which suggests that the gas falls quickly into the center of the halo from large distance in a time comparable to the halo free-fall time.
The clumps that are formed {\sl in situ} migrate rapidly inwards while releasing their residual angular momentum to the stellar component  surrounding the bulge \citep{immelietal04, elmegreenetal08, bournaudetal11}.

\subsection{Inflow through the disc {\it versus}
 direct CGM infall}

\begin{figure*}
   \includegraphics[width=0.65\columnwidth]{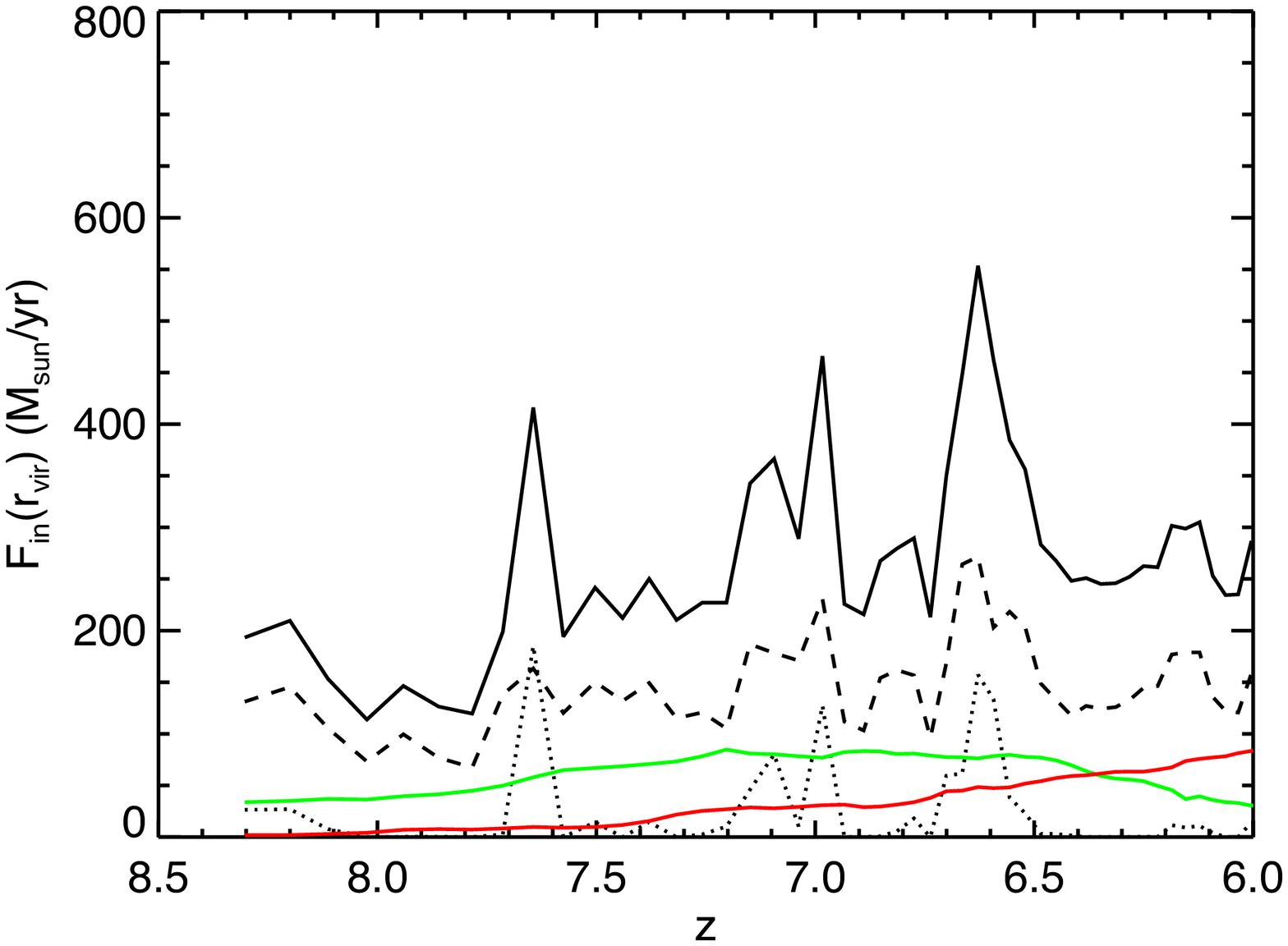}
   \includegraphics[width=0.65\columnwidth]{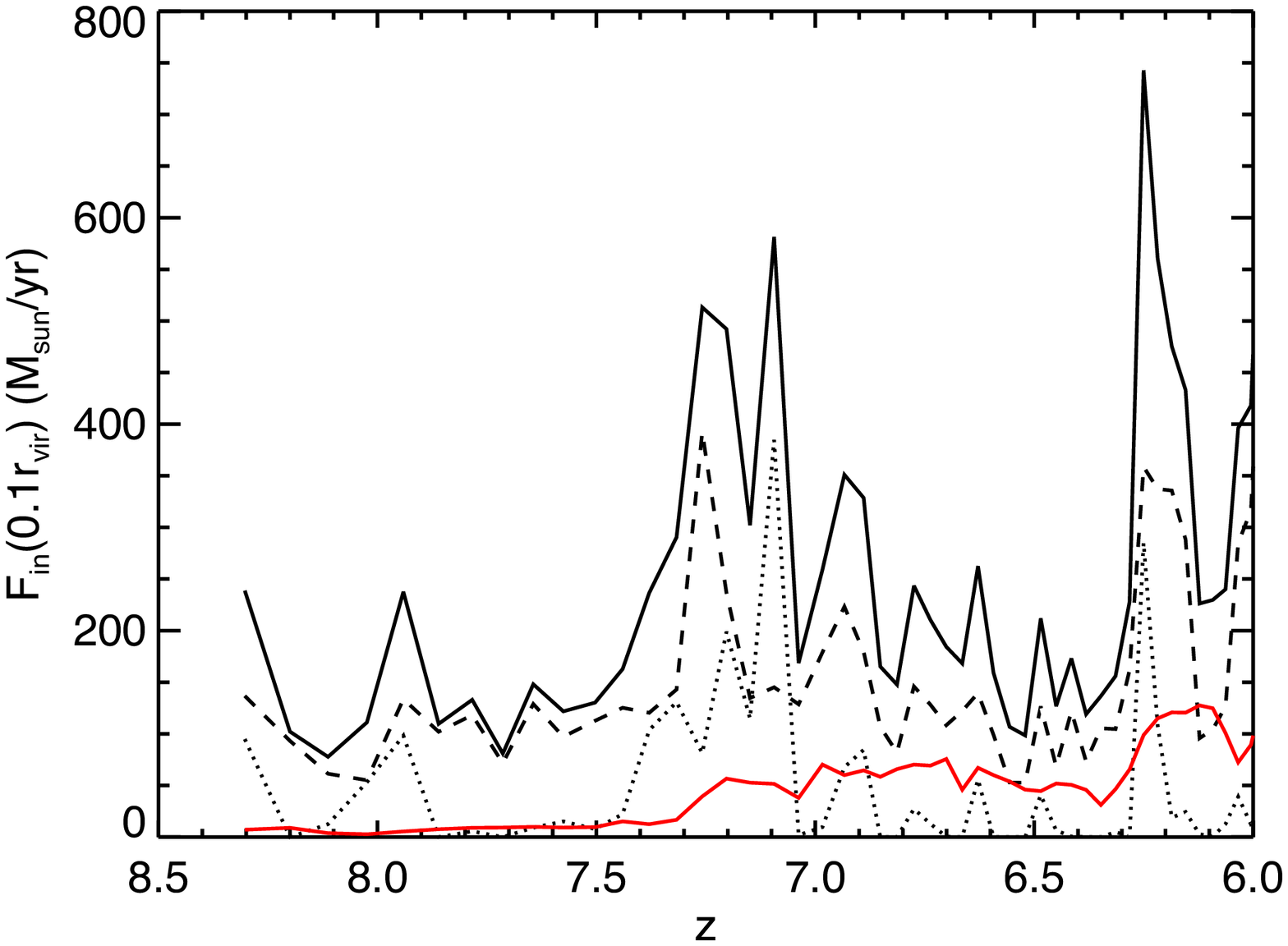}
   \includegraphics[width=0.65\columnwidth]{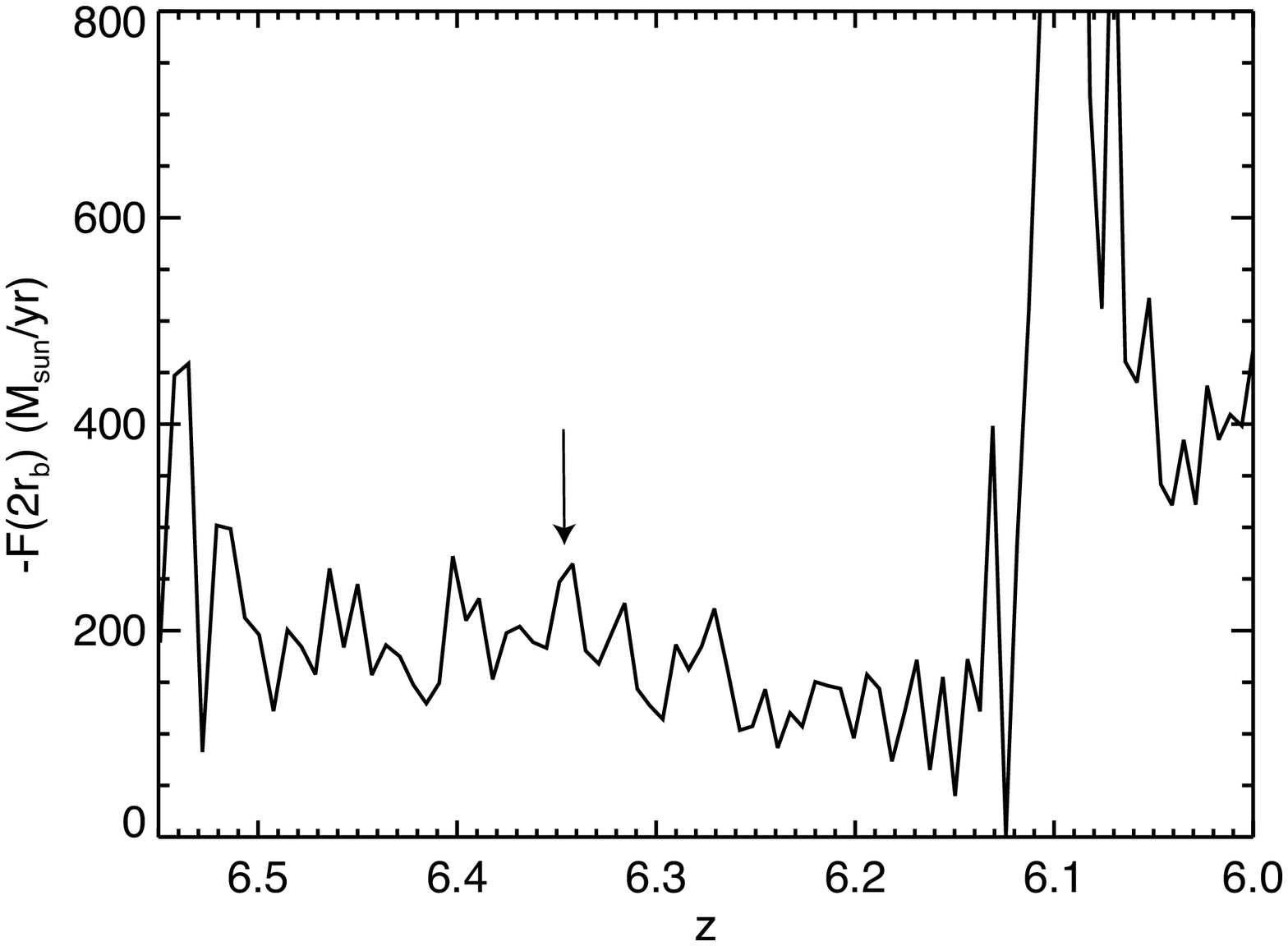}
      \caption{{\sl Left panel}: inward flux measured at $r_{\rm vir}$ for the SHhr simulation for the star-forming gas (dotted line), the cold filamentary gas (dashed line), the cold diffuse gas (green line), the hot shock-heated gas (red line), and the total (solid line). {\sl Middle panel}: same as the left panel measured at $0.1\, r_{\rm vir}$. The gas is essentially accreted through cold infall (diffuse and filamentary) at $r_{\rm vir}$, and through cold filaments at $0.1\, r_{\rm vir}$, with very few star-forming clumps contributing to the overall mass infall. {\sl Right panel}: net inward flux ($-F=-(F_{\rm in}+F_{\rm out})$) measured at twice the bulge radius $2 r_{\rm b}=80$ pc. The gas entering the bulge of the galaxy is composed of star-forming clumps with a bursty accretion of gas. The vertical arrow marks the burst corresponding to the accretion event seen in fig.~\ref{fig:clump_migration}. The massive burst of accretion at $z=6.1$ corresponds to the capture of a satellite galaxy.}
\label{fig:fluxSHhr}
\end{figure*}
%
\begin{figure*}
   \includegraphics[width=0.333\columnwidth]{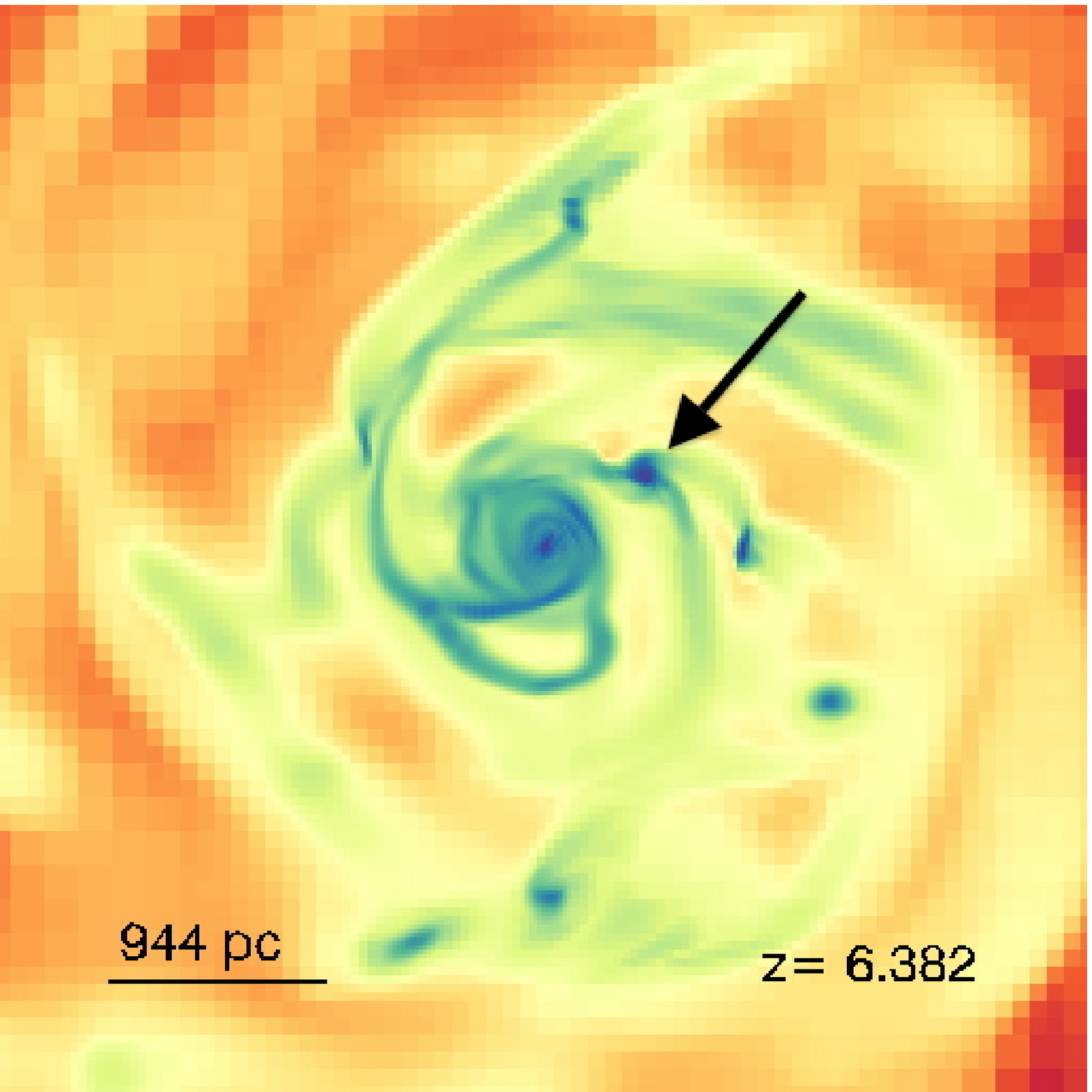}
   \includegraphics[width=0.333\columnwidth]{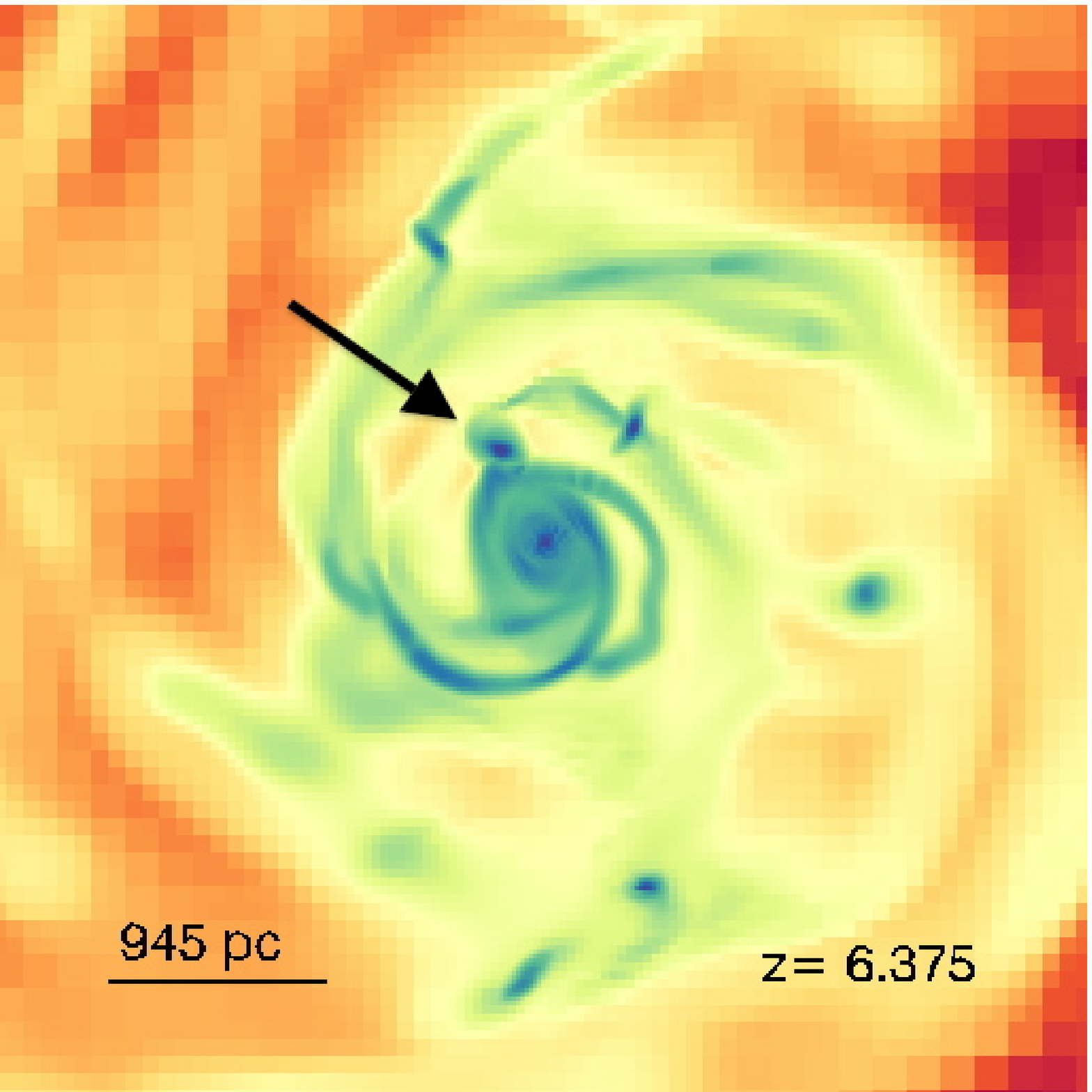}
   \includegraphics[width=0.333\columnwidth]{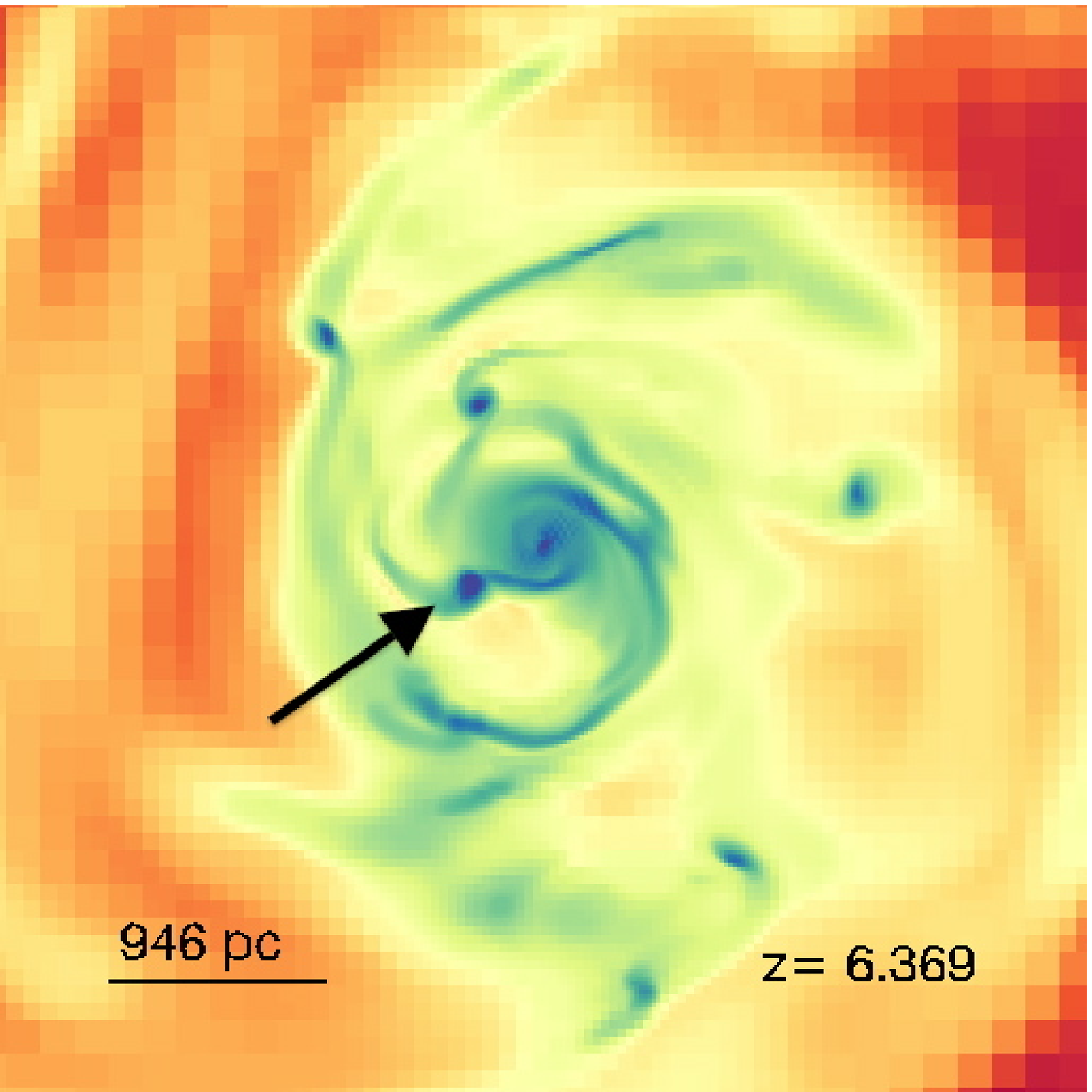}
   \includegraphics[width=0.333\columnwidth]{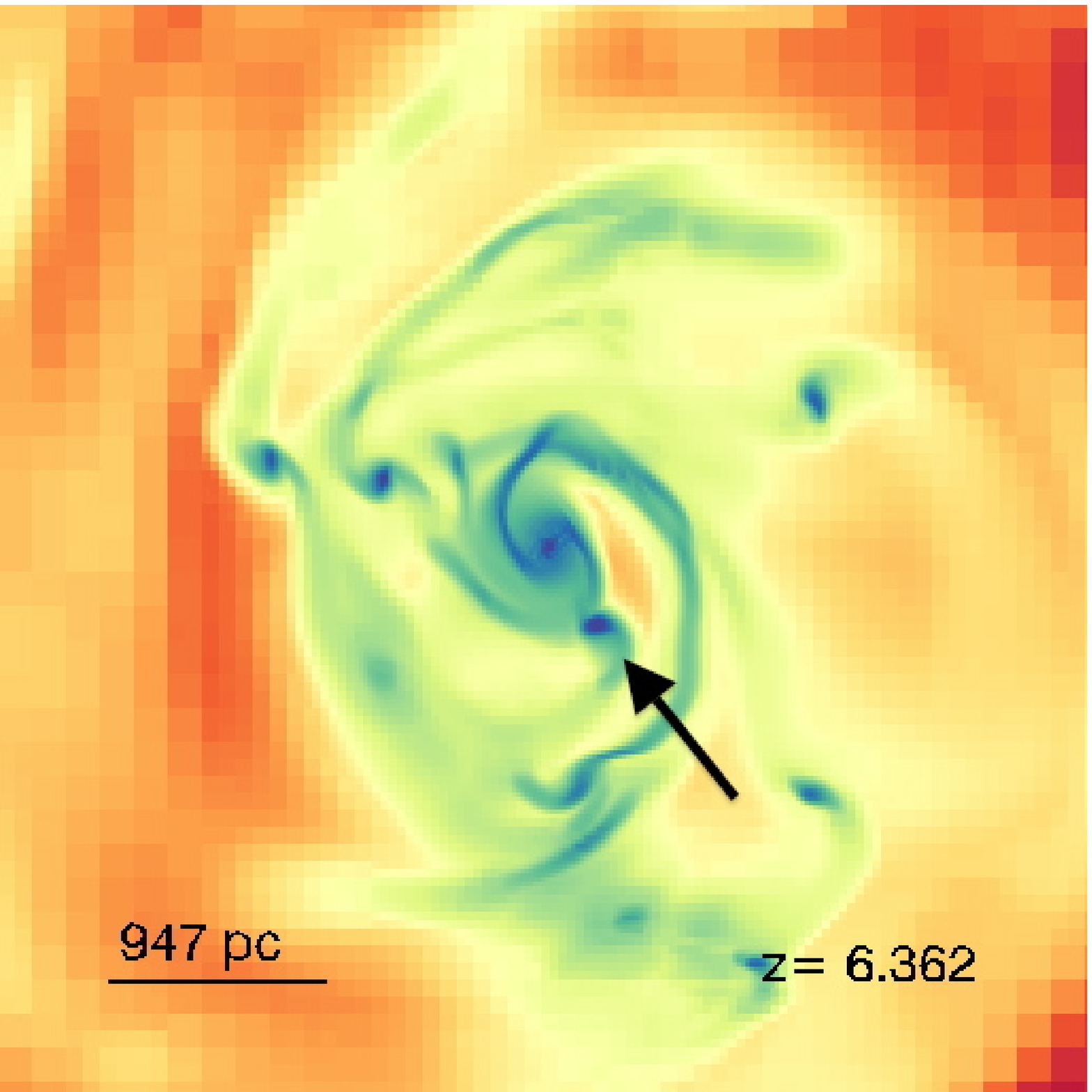}
   \includegraphics[width=0.333\columnwidth]{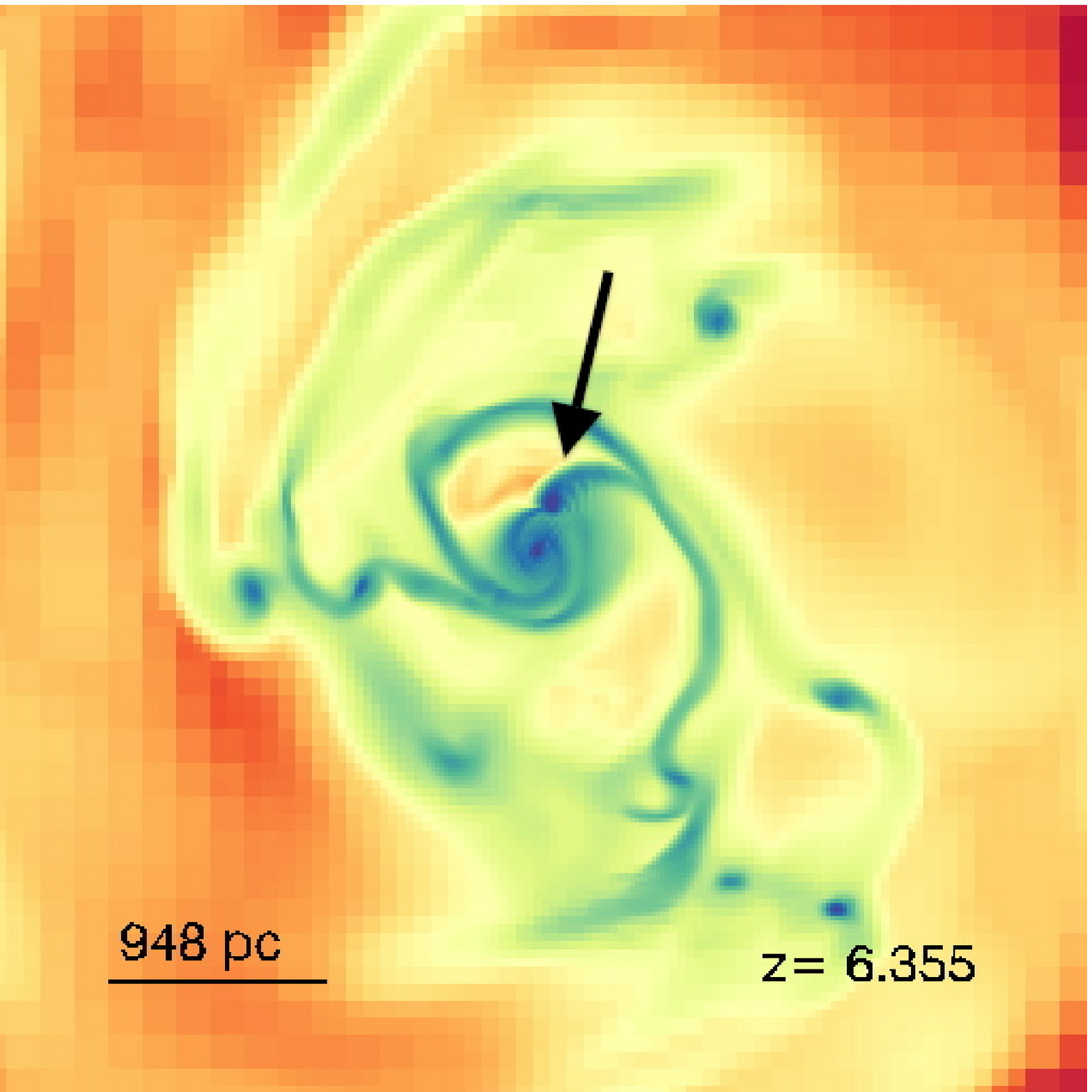}
   \includegraphics[width=0.333\columnwidth]{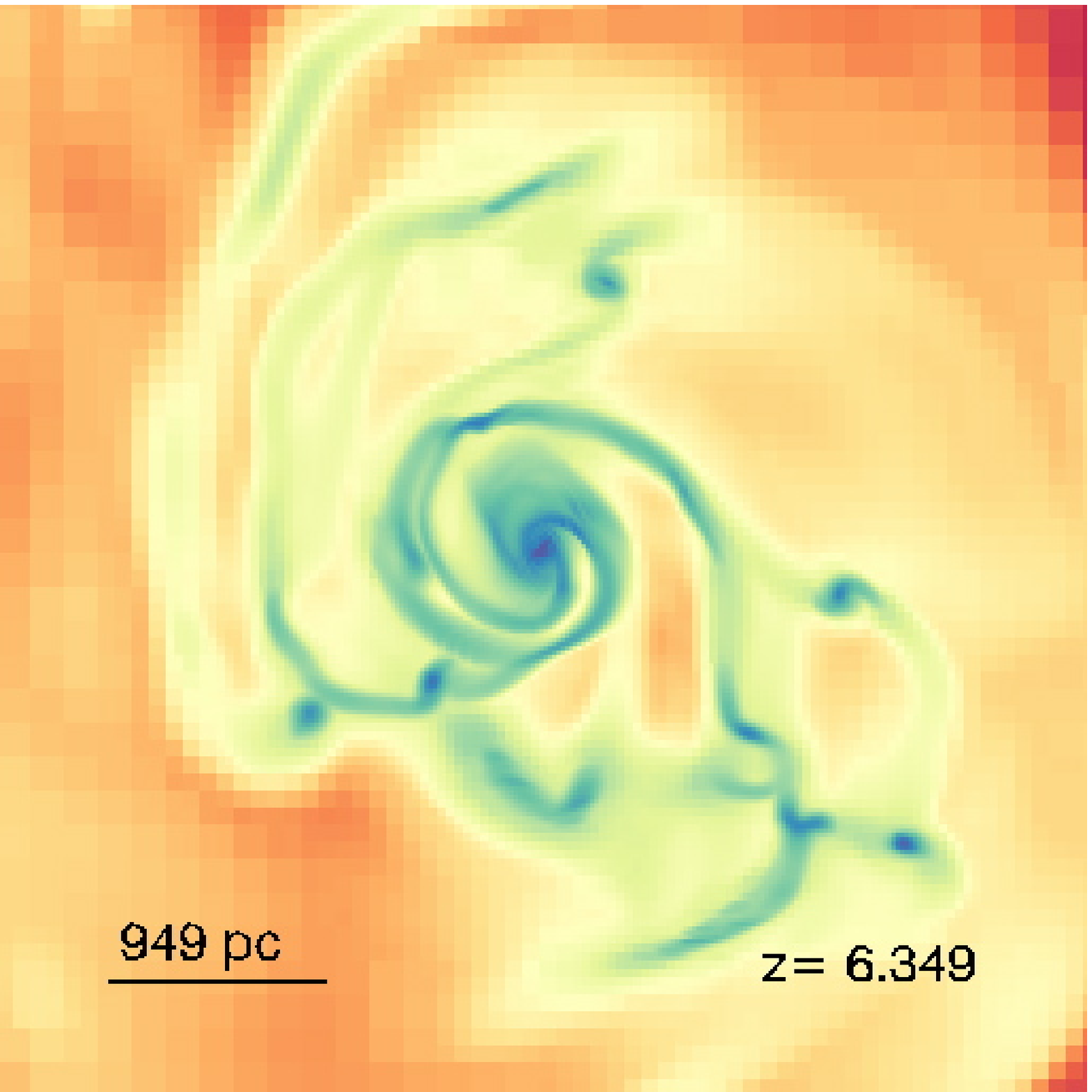}
      \caption{Time sequence ({\sl from left to right}) of the gas density evolution in the central SHhr galaxy illustrating the migration of a dense clump of gas (indicated by the black arrow) into the central bulge. This radial migration of {\sl in situ} clumps is consistent with the inflow burst of figure \ref{fig:fluxSHhr} (right panel, vertical arrow).}
\label{fig:clump_migration}
\end{figure*}

We now investigate how much gas  flows  directly from the CGM  into the bulge
and how much gas passes through  the central disc surrounding the
bulge first.  We define disc\footnote{Note that this definition differs from that given in fig.~\ref{fig:momemtumvsz}.} tracer particles that have passed through the disc
surrounding the bulge before settling into the bulge as those that  have stayed
at a distance $r_{\rm b} < r < 0.1\,  r_{\rm vir}$ (where the scale length of the bulge $r_{\rm b}=200$ pc for the SH and LH runs, and $r_{\rm b}=40$ pc for the SHhr run) for a time $t_{\rm acc}$ larger than the time
required to complete a full rotation  at $0.1 r_{\rm vir}$,
$t_{\rm circ} (0.1\, r_{\rm vir})=2\pi (0.1 \, r_{\rm vir})/v_{\rm circ}(0.1\, r_{\rm vir})$, with $v_{\rm circ}(0.1\, r_{\rm vir})$ being computed from the mass of gas, stars, and DM at $0.1 r_{\rm vir}$. 
Looking again at the selection of tracer  particles shown in 
fig.~\ref{fig:disc_cgm_selection}, most of the bulge tracer particles 
accreted from the CGM fall very rapidly into the bulge, with a few
that appear to have their trajectories perturbed during their infall but \emph{before}
reaching the disc ($0.1 \,  r_{\rm vir}$). Some other tracer particles
hang around in the disc for some time \emph{after} reaching $0.1 \, r_{\rm vir}$ 
at an almost constant distance from the centre before being accreted into the
bulge. These latter particles are the ones which migrate from the disc to the bulge.

\begin{figure*}
   \includegraphics[width=0.66\columnwidth]{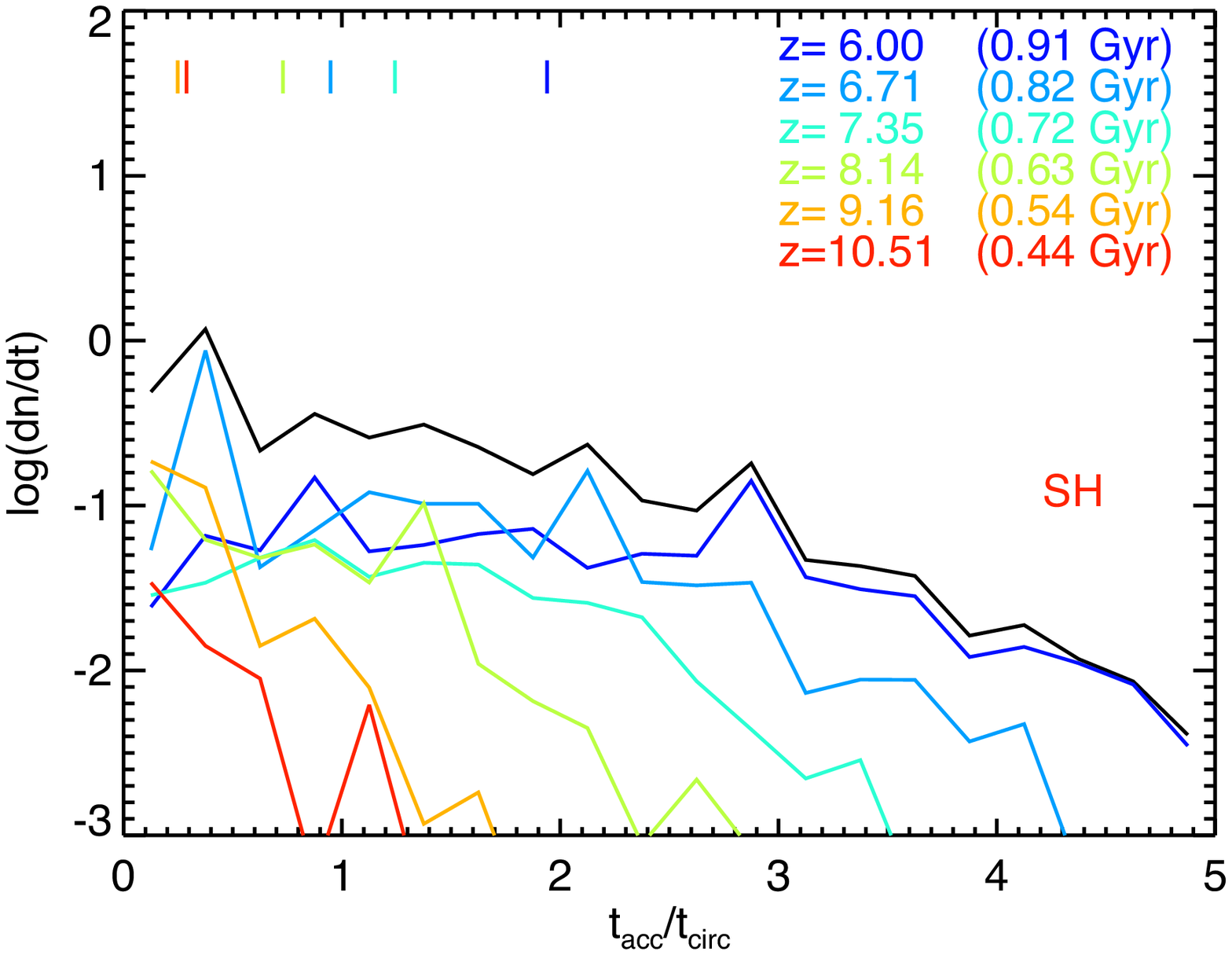}
  \includegraphics[width=0.66\columnwidth]{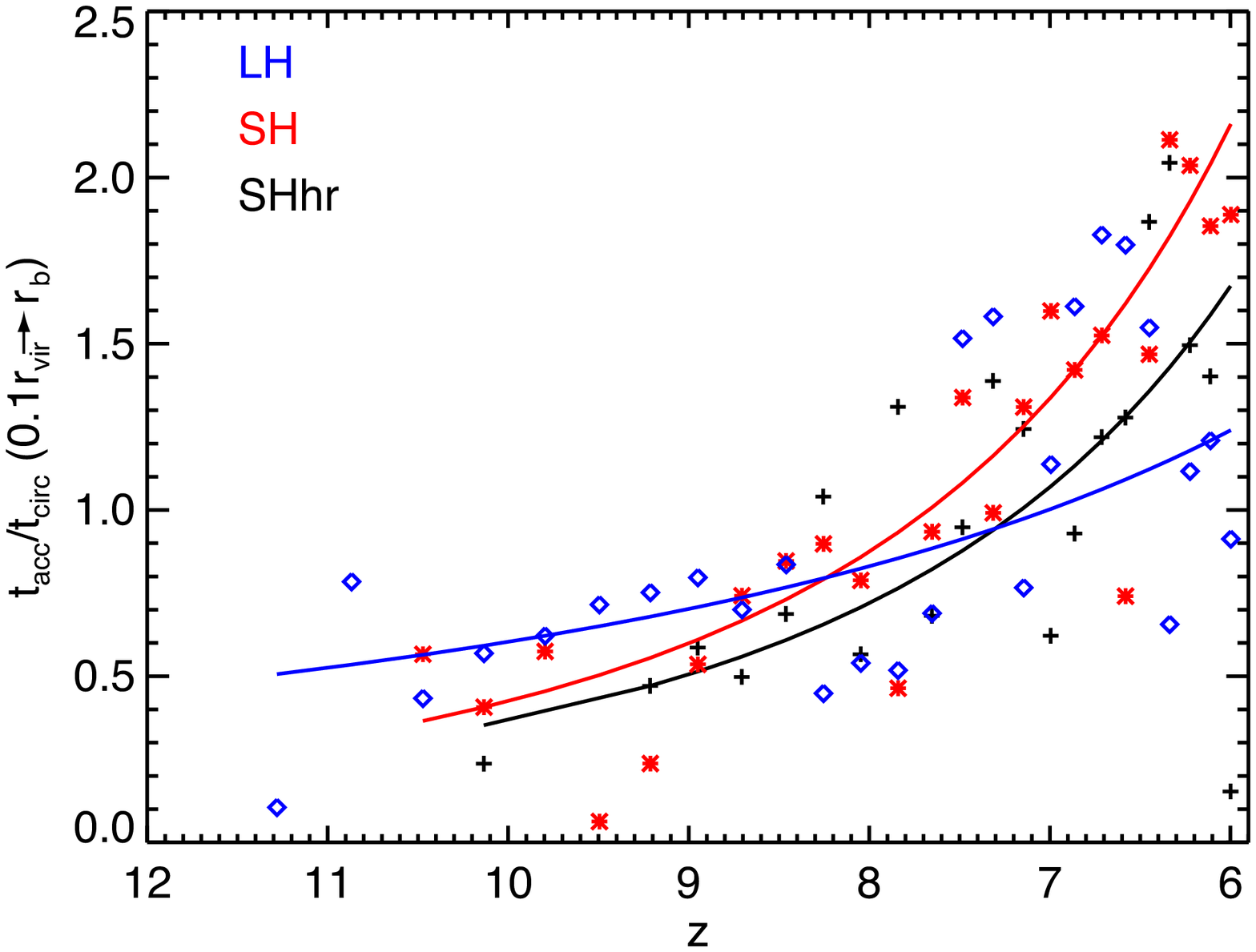}
  \includegraphics[width=0.66\columnwidth]{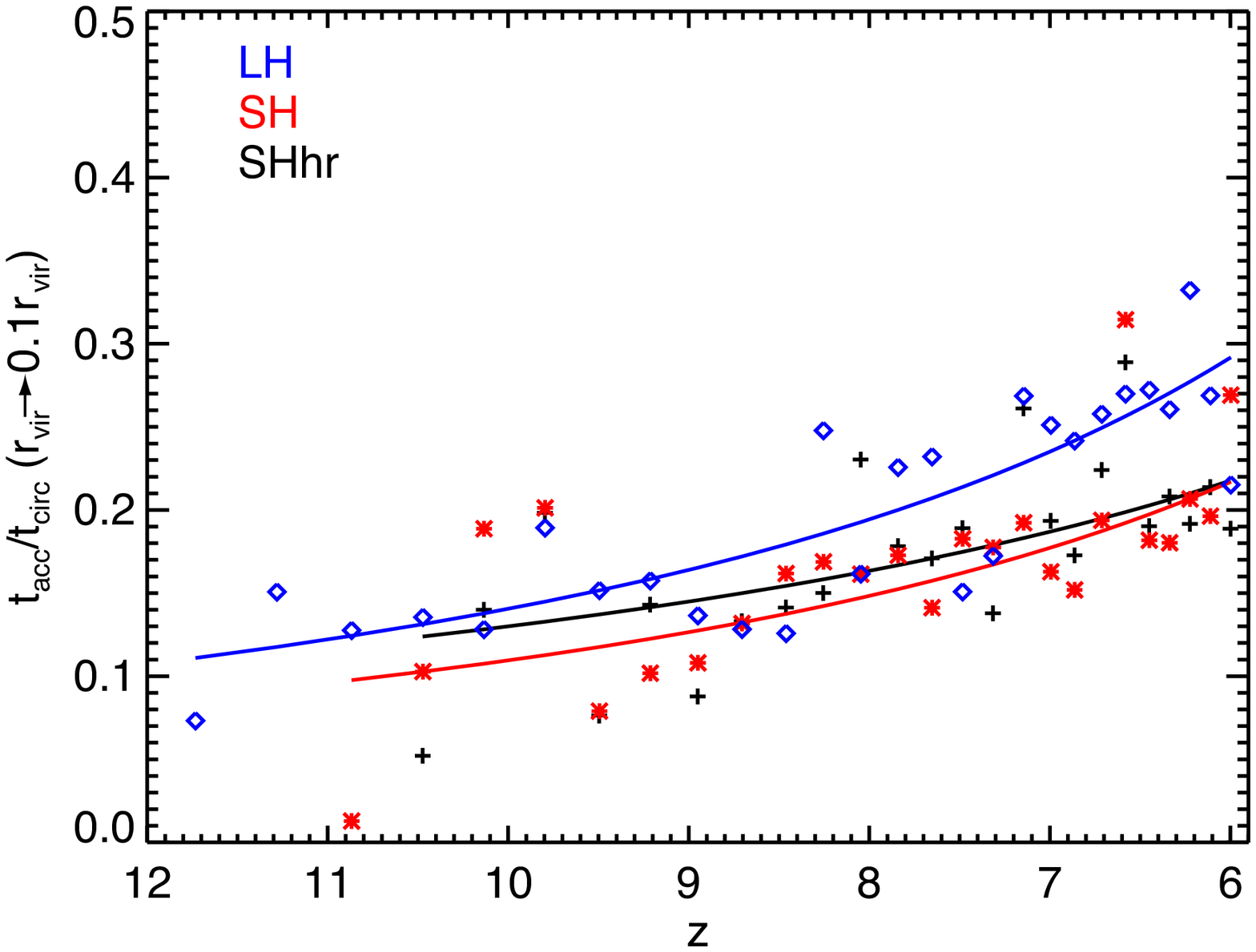}
       \caption{ {\sl Left panel:}  The number of  tracer
         particles identified in the bulge at $z=6$ in the SH
         simulation   as a function of the time (normalized to $t_{\rm circ}(0.1 r_{\rm vir}$) 
     spent at radii  $r< 0.1 \, r_{\rm vir})$
for different accretion redshifts as color coded on the plot. The
vertical ticks correspond to the average accretion  time
for tracer particles entering the bulge at the  corresponding
redshift.   {\sl Middle panel:} average normalized accretion time of bulge tracer particles at $z=6$ to travel from  $0.1\, r_{\rm vir}$ to $r_{\rm b}$. The number of completed orbits  spent in the disc is still rather small (less than 2).  As expected, the later the infall, the larger the number of completed orbits. 
{\sl Right panel:} same as middle panel but to travel from $r_{\rm vir}$ to $0.1\, r_{\rm vir}$ for the three different simulations, LH (blue diamonds), SH (red stars), and SHhr (black plus). Particles travel in the halo very quickly in less than a fourth of a rotation (i.e. in a free-fall time).
}
\label{fig:histo_distance_accretion1}
\label{fig:time_fit}
\end{figure*}

This is further quantified in fig.~\ref{fig:histo_distance_accretion1}
which  shows  number counts of tracer particles identified as
belonging to the bulge at $z=6$. The figure shows how much time
$t_{\rm acc}$ (as a function of $t_{\rm circ}$) the tracer particles spend at radii $r_{\rm b} < r <
0.1\,  r_{\rm vir}$   as a function of the time before they reach $r_{\rm b}$. 
The time to complete  one rotation at $0.1\, r_{\rm vir}$, $t_{\rm circ}$, increases with time, thus, the increase seen in fig.~\ref{fig:histo_distance_accretion1} corresponds to a net increase of accretion time for baryons before being accreted into the bulge.
Tracer particles which are accreted earlier fall rapidly into the bulge, while
particles accreted later spend more time within the disc. Indeed, the
particles accreted later have typically a larger angular momentum
modulus \citep{pichonetal11}, hence are expected to settle into the  disc 
before they reach the bulge. This pattern  is the same for all three simulations: 
for any given patch of material, late-time infalling material  
does so with increasing angular momentum. 

We now measure the evolution of the average relative accretion time
with redshift $t_{\rm acc}/t_{\rm circ}=\tau (1+z)^\alpha$ for our
three different simulations. We find $\tau=$ 3.4, 3.1, 1.4 and
$\alpha=$ $-3.6$, $-3.4$, and $-1.6$ for the SH, SHhr, and LH halos
respectively. In  the most massive LH halo, we clearly  have
shorter relative accretion time compared to the lower mass SH halo
(middle panel of fig.~\ref{fig:time_fit}), suggesting that the bulge in
this most massive structure acquires more gas by direct
infall from the CGM. For the smaller mass halo, in the high resolution
simulation SHhr, the particles have slightly shorter accretion times than the
low resolution simulation SH. The SHhr
simulation  is able to follow the formation of smaller 
clumps which have shorter accretion times than matter falling in
diffusely. This is in agreement with clumps in the galaxy having shorter migration times
towards the bulge than gas being accreted by secular disc
instabilities such as bars,  or spiral arms \citep{bournaudetal05, bournaudetal11}.

If we similarly consider the time it takes for these particles to travel from the virial radius, $\, r_{\rm vir}$,  to the disc, $0.1\, r_{\rm vir}$, it becomes apparent  that these same particles on average never complete one full rotation corresponding to $t_{\rm circ}(r_{\rm vir})$  ($\tau=$ 0.6, 0.3, 0.8 and $\alpha=$ $-1.5$, $-1.1$, and $-1.6$ for the SH, SHhr, and LH halos, see right panel of fig.~\ref{fig:histo_distance_accretion1}).
In fact, it takes them  a time comparable to the halo free-fall time ($t_{\rm ff}$=$t_{\rm circ}/\sqrt{32}$) to reach the galaxy.   For these massive halos at high redshift, the gas that makes up  the compact bulges is brought in from the IGM almost directly.  This is in good agreement with our previous finding that the accretion rate at $0.1\, r_{\rm vir}$ is comparable to the accretion rate at $r_{\rm vir}$ (see fig.~\ref{fig:fluxSHhr}).

\begin{figure*}
  \includegraphics[width=0.93\columnwidth]{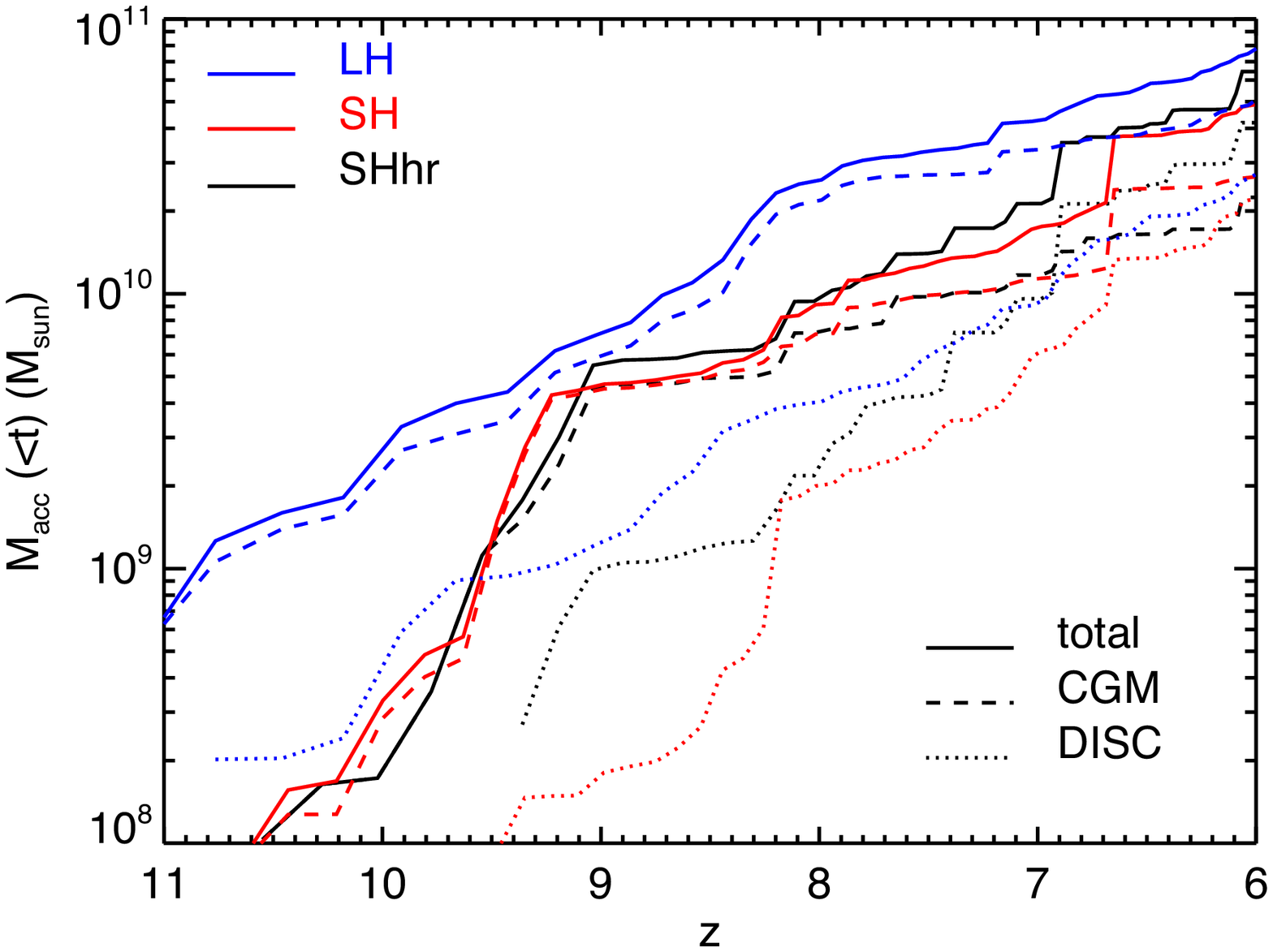}
  \includegraphics[width=\columnwidth]{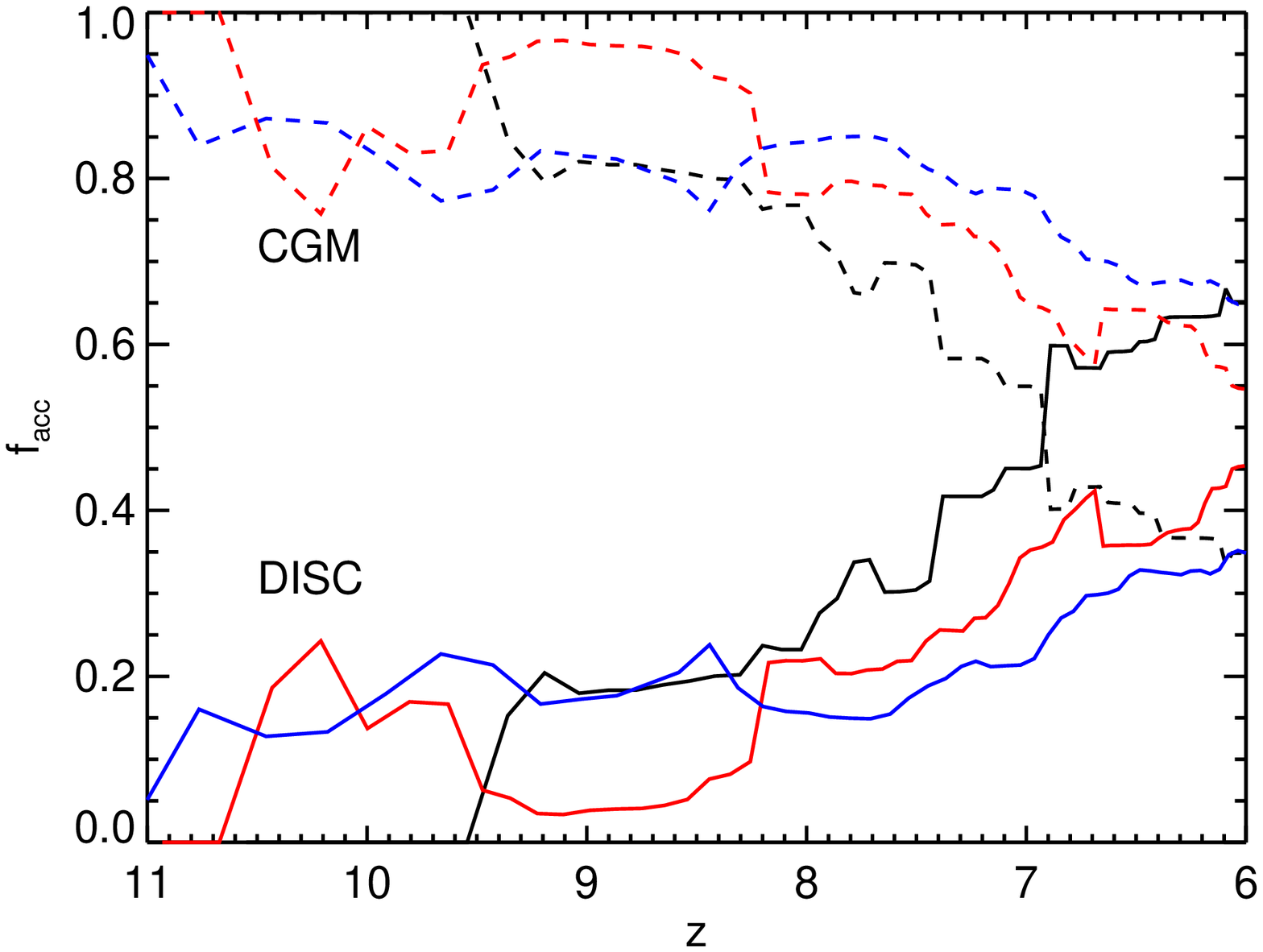}
      \caption{ {\sl Left panel:} Cumulative mass of gas (tracer particles) accreted into the 
      bulge as a 
        function of cosmic time, for gas directly accreted from the CGM (dashed curves)
       and gas accreted through a disc instability (dotted curves). Blue curves  correspond to the SH,  red curves  to 
        the LH and  black curves  to  the SHhr  simulations.
         {\sl Right panel:}  Fraction of tracer particles accreted directly from the CGM (dashed curves) or  through the 
         disc  (solid curves). The color coding is the same as for the left panel.
         At high redshift infall is predominantly directly  from the CGM, while by redshift $z=6$,  about half  of the 
         particles spend more than one $t_{\rm circ} (0.1 r_{\rm vir}) $ in  the disc.
  }
\label{fig:total_accretion_disc_cgm}
\end{figure*}

We can now define gas as flowing directly into the bulge from
the CGM as gas for which the time to reach the bulge from $0.1\,r_{\rm vir}$ is smaller than the time to 
make a full revolution at $0.1\,r_{\rm vir}$ {\it i.e.} $t_{\rm acc}<t_{\rm circ}$. Correspondingly
for gas flowing into the bulge through the disc driven inwards by 
disc instabilities, $t_{\rm acc} \ge t_{\rm circ}$. Figure~\ref{fig:total_accretion_disc_cgm}
shows that the fraction of baryons accreted directly from the CGM into
the bulge defined in this way is dominant early on and decreases to about half
by $z=6$.  Disc instabilities are increasingly important 
for the feeding of the bulge component of the galaxy at late times. 
This is in good  qualitative agreement with
the prediction of~\cite{pichonetal11}: early infalling gas has very
low angular momentum, and can directly supply the fuel for the growth
of the massive compact bulge, gas accreted at late times coming from larger
distance in the IGM and with higher angular momentum contributes to
the formation of a compact and gravitationally unstable disc. 
This is consistent with the disc size evolution with time (increased angular momentum), that grows from 100 pc at $z=9$ to 350 pc at $z=6$ in the SHhr simulation.
The total baryonic mass accreted into the bulge, as indicated by the amount of tracer particles accumulated into this compact region, is larger than the total mass of gas plus stars measured at $z=6$ into the bulge by a factor $1.5-3$ (table~\ref{tab:feature}). As the tracer particles do not account for the amount of gas mass lost to star formation and that stars can be expelled from the compact bulge by mergers, tidal stripping, etc., more than the gas we systematically over-estimate the amount of baryons ending up in the bulge with this method.

Still, there clearly is a sufficient amount of gas in the bulge at all times to potentially feed a very massive BH with mass $\sim10^9\, \rm M_\odot$ (see numbers in table~\ref{tab:feature}).
The BH mass to bulge mass ratio $\Gamma=M_{\rmÊBH}/M_{\rm bulge}$ evolve with redshift as seen in observations \citep{decarlietal10, merlonietal10}  and predicted by cosmological simulations \citep{dimatteoetal08, booth&schaye11, duboisetal12}.
With the extrapolation of the observational fit from \cite{decarlietal10} (obtained between $0<z<3$) up to $z=6$ ($\Gamma=0.096$), we expect the BH mass to be $M_{\rm BH}\sim 2 \cdot 10^9 \rm \, M_\odot$. 
The amount of gas in the bulge of the SHhr galaxy at $z=6$ could seem insufficient to feed a BH as massive as $2\cdot 10^9 \rm \, M_\odot$.
However, the accretion and consumption of gas onto the BH is in competition with the consumption of gas through the star formation process, and a non-negligible amount of gas that is turned into stars could be captured by the central BH if there was any.

When measuring the gas mass flux around the galactic bulge in the SHhr galaxy (right panel of fig.~\ref{fig:fluxSHhr}), it becomes evident that the accretion is very irregular and proceeds by bursts of very dense clumps of gas, with a high level of $\sim200 \, \rm M_\odot/yr$ gas accretion rate  towards the bulge maintained.
Such a mode of accretion onto the bulge is illustrated in fig.~\ref{fig:clump_migration}, where we can observe a dense clump of star-forming gas migrating into the center of the galaxy and being absorbed by the bulge.
From this stochastic accretion mode, we expect the central AGN to release energy in rapid bursts interspersed with longer periods of very little activity.
Capture of galaxy satellites (burst of accretion at $z=6.1$) are rarer events that bring huge amounts of gas ($>1000\, \rm M_\odot/yr$) and can potentially disperse all the galactic gas while triggering a strong AGN activity \citep{dimatteoetal05}.

\subsection{How the gas loses angular momentum to form bulges}

If we look back at the angular momentum evolution plotted in 
fig.~\ref{fig:momemtumvsz}, we realise that  bulge tracer particles 
sample material which has a factor $\sim 3$  less specific angular momentum
to begin with in the LH run than the average specific angular momentum of the baryons in the halo. 
This is obviously a segregation effect.  Amongst all the material flowing into
the halo, it is the early infalling low angular momentum  gas that
preferentially settles (and much of this directly from the CGM) into
the bulge as  discussed extensively in the last section and in
\cite{pichonetal11}.  This is, however, not quite sufficient to explain the 
factor 10-30 lower angular momentum of the gas and stars in the bulge. 
The right panel of  fig.~\ref{fig:disc_cgm_selection} 
shows  how  the angular momentum  of the ensemble of tracer particles 
which we have discussed earlier lose their angular momentum as they sink 
to the central bulge. The angular momentum loss is complex with some of the particles 
losing  it rather gradually, but there are also sudden drops by factors of a few, indicative of what we may call angular 
momentum cancellation, followed by smaller gains of specific angular momentum. 

\begin{figure*}
   \includegraphics[width=\columnwidth]{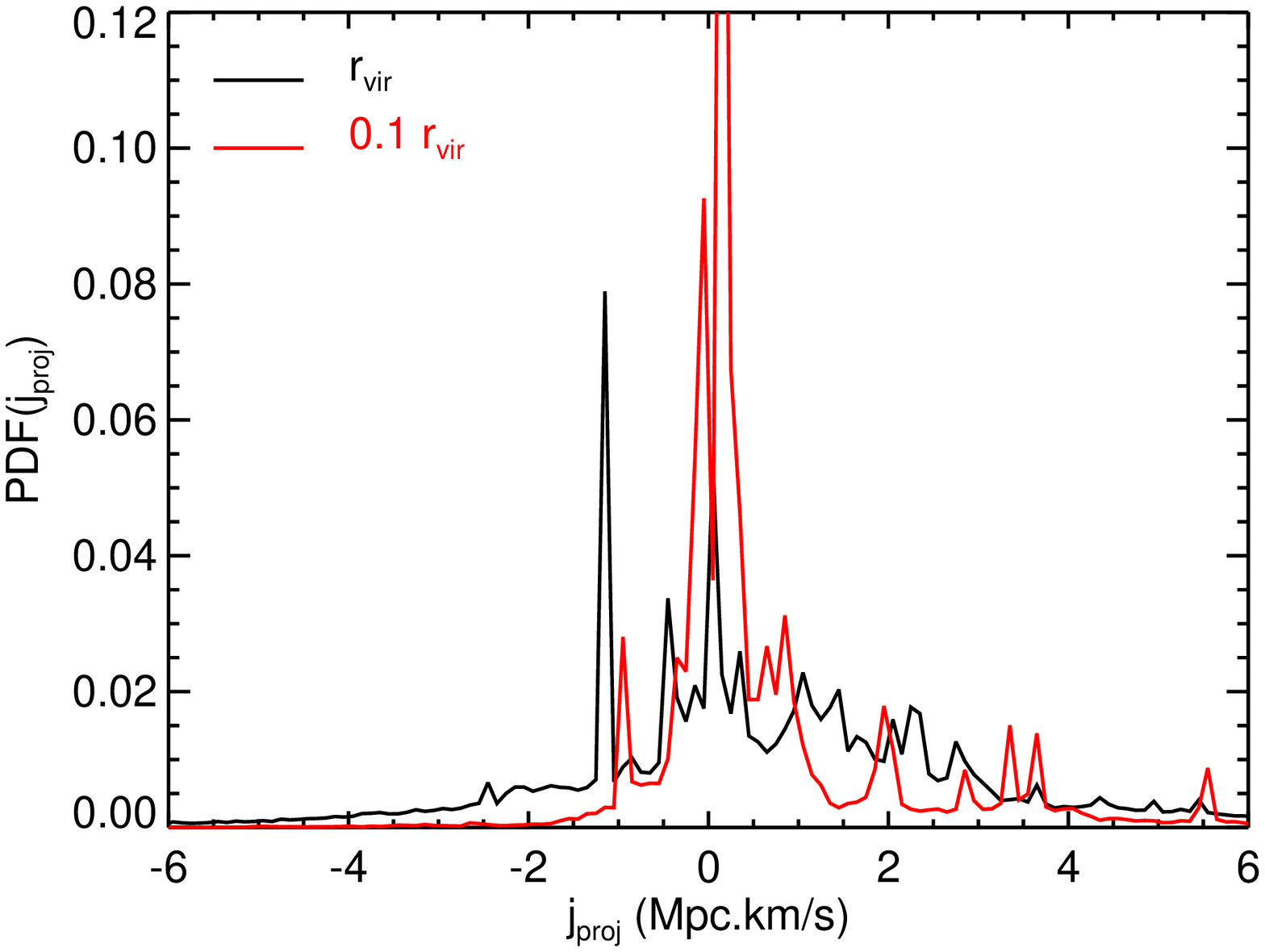}
    \includegraphics[width=\columnwidth]{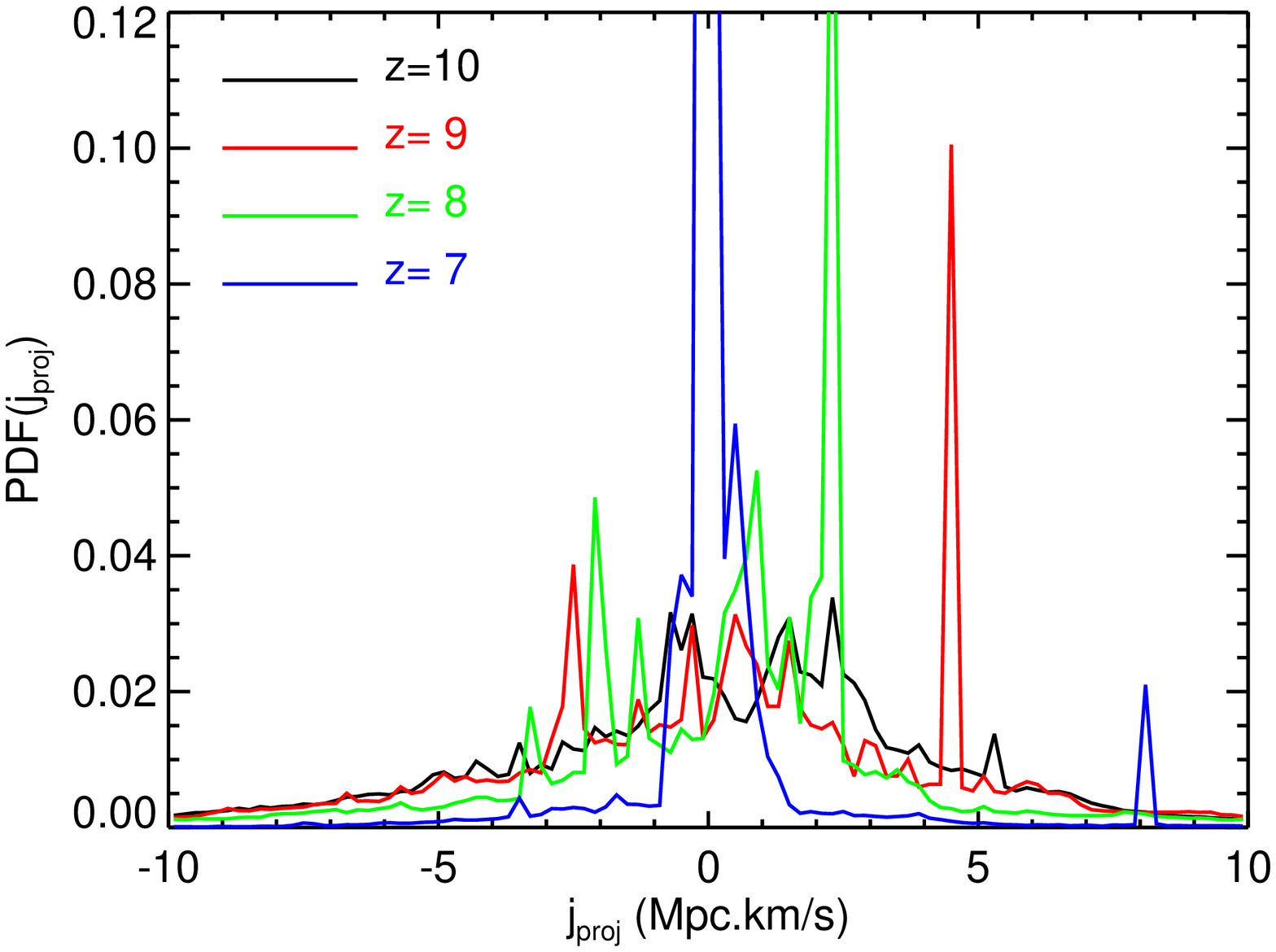}
      \caption{{\sl Left panel: } PDF of the projected specific angular momentum of bulge tracer particles of the SH simulation selected at $z=6$ at the virial radius (black line) and at $0.1 \, r_{\rm vir}$ (red line). The specific angular momentum of each particle is projected along the normalized specific angular momentum of the patch containing all tracer particles
      in each region. The specific angular momentum is computed relative to the centre of the bulge. Note the overall contraction
     of the angular momentum distribution towards the centre of the halo. 
{\sl Right panel:} PDF of the projected specific angular momentum of the tracer particles of the bulge at $z=6$ along the angular momentum of the patch for different redshifts $z=10$ (black line), $z=9$ (red line), $z=8$ (green line), and $z=7$ (blue line). The specific angular momentum is computed relative to the centre of mass of the patch of bulge tracer particles. Note again the contraction of the accumulated angular momentum PDF as a function of cosmic time.}
\label{fig:pdf_jproj}
\end{figure*}

In the left panel of fig.~\ref{fig:pdf_jproj}  
we compare  the PDF of the projected angular momentum  of bulge tracer particles 
at $0.1\, r_{\rm vir}$ at $z=6$ with the angular momentum of tracer particles at $r_{\rm vir}$ \citep{vdboschetal02,sharmasteinmetz05}. 

As discussed by  \cite{kimmetal11}, if the gas is allowed to cool, this specific angular momentum distribution becomes very broad. 
A significant fraction of the baryons is actually rotating in the
direction opposite to the overall rotation of the halo. Filaments and clumps of material with very
different angular momentum orientation will reach the 
centre in turn, with low specific angular momentum material getting there first. 
This explains why in  the right hand panels of 
fig.~\ref{fig:healpix_map}  the angular distribution 
of the mass influx changes rapidly with time deep inside the 
halo. The figure clearly demonstrates that the randomly oriented angular momentum 
of the material  that reaches the bulge  cancels each
other out resulting in a steep spike around zero in the PDF of the angular momentum. 
This together with the usual means of losing angular
momentum like the transfer between spin and orbital angular momentum,
dynamical friction acting on in-spiraling clumps and  gravitational 
 instabilities (bar-like and clumping) in the disc  results in a very compact but nevertheless 
only moderately rotating bulge.   Note that a  similar "cancellation" of angular momentum is seen 
as the tracer particles approach the halo (right panel of  fig.~\ref{fig:pdf_jproj}). 

\subsection{What is different in rare halos?}

{\it Why is this angular momentum
segregation  so  efficient in the two massive rare
halos we have studied here?} To answer this properly would require 
a systematic study of a wider range of halo masses and redshifts.
Here we formulate a brief argument based
on a comparison with the previous work \cite{kimmetal11} and \cite{pichonetal11}.   

Expanding locally on the work of \cite{bkp96}, these authors   showed  that 
 the large scale distribution of the angular momentum
 in a collapsing patch produces a consistent  secondary inflow:
  gas is  draining out of  the prominent surrounding  voids in the cosmic web, into the
 sheets  and filaments before it finally gets accreted onto virialised
  halos along cold flows. As these sheets/filaments constitute the boundaries
 of  generally asymmetric voids, they acquire a net transverse motion relative to their host halo, which becomes the seed of the advected angular momentum along filaments.
 They argued that this large-scale driven consistency  explains why cold flows
  are so efficient at building up  discs from the inside out.
  \cite{kimmetal11} and \cite{pichonetal11}   also
 reported using the MareNostrum hydrodynamical simulations   \citep{ocvirketal08}  that, unlike that of DM\footnote{though \cite{knebe&power08}  report a small trend of spin parameter decreasing with halo mass at high redshift}, the spin parameter
 of the baryons decreases (i) as the halo mass increases and (ii) as redshift increases 
for a halo of fixed mass  (in their  fig.~4  for the angular momentum flux
of the gas at the virial radius, and in fig.~8 of \cite{kimmetal11} for the specific angular momentum of the gas  
within $0.1 \, r_{\rm vir}< r<r_{\rm vir}$).  This suggests
a decrease of the overall gas spin parameter for rarer (and
thus more massive halos).  

Therefore we  now   argue that with the
formation of the very compact bulges in the extremely rare halos
that we studied, we see a continuation of this trend which feeds through to 
the angular momentum segregation and cancellation during the
non-linear evolution (``collapse'') of the halo.  In other words, 
the gas  streaming cold to the centre avoids significant mixing in the outskirts of the dark halo.
 It thus inherits  the  decrease of  spin parameter of 
 the gas with increasing rareness of the density  peak seeding the formation of the halo.
 
This trend is expected on theoretical grounds.
   Indeed, \cite{Pichonetal10}, (fig. 4, bottom panel) demonstrates that in Lagrangian space (corresponding to the large redshift Gaussian initial conditions for structure formation),  the rarer the peak, the larger the number  of filaments connected to it (as the high contrast requires large negative curvature in all directions, which implies isotropy of the peak;  this isotropy allows for connections to filaments from all directions). 
Hence we would expect that the larger number of connected filaments lead to more 
efficient cancellation of their specific angular momentum 
resulting in a low overall spin for the bulge. Part of this cancellation already occurs on large 
scales  in the cosmic web during shell crossing of the DM in the cosmic walls/sheets 
and filaments, and continues when the gas with  the low angular momentum advected 
along the filamentary streams reaches the bulge at the  centre of the halo.   
Note that tidal torque theory also predicts a weak  anti-correlation \citep{heavens&peacock88}
between the spin parameter and the mass of halos, as the amount of torquing 
between a rare  density peak and its (more spherical) tidal field   decreases  with the 
rareness of the peak.  For the baryons, this anti-correlation is amplified by the shock-induced segregation
which takes place in the cosmic web (walls, filaments).

Let us revisit the early stage of the bulge formation in our simulations in view of these theoretical priors.
First, the  low angular momentum cold gas streams right  
 along a couple of filaments into the seed of a bulge. 
Gas falling in later  along other filaments is
unlikely to have low {angular } momentum with respect to that  frame 
(and typically will have larger angular momentum amplitude) 
so will recurrently form a transient disc. This disc gets disturbed by minor mergers which also sink-in via
angular momentum loss with the stellar component at these radii. The competing effects  of disc formation and disc disruption  depend on the rareness of the peak as well as cosmic time.
The rarer the peak, the lower the spin parameter of the advected gas.  
The later the infall of the gas, the larger  the amount of  advected angular momentum.
The two halos in our SH and LH simulations  are rather extreme examples
in terms of rareness and appear to confirm this picture. Hence
the formation of the very compact galactic bulges 
in the two simulations corresponds to the early stage where low angular 
momentum gas streams directly to the bulge  and the rotationally supported
 disc is small and constantly destroyed and recreated.
 The later stages, when higher angular momentum gas accumulates and settles into a robust disk as described in  \cite{pichonetal11}, may occur later  depending on the nature of the Lagrangian peak.
For very rare massive  halos such as those simulated here,  the opportunity to reach this second stage has 
not occurred by redshift $z=6$.
   
\section{Discussion}
\label{sec:discussion}

\subsection{Limitations and Caveats}

We would like to stress again  that we have performed in this paper an exploratory study which consists in the analysis of high resolution simulations of only two massive halos.
These simulations  \emph{do not} include any model for the growth of a central BH, and perhaps more importantly neither AGN nor stellar feedback.
We have mainly  concentrated here on describing some of the physical  mechanism of the inflow of gas in these massive halos with a particular emphasis on the role played by angular momentum. 
AGN (and stellar) feedback  together with the effect of changing the details of star formation in the simulations  are  probably the most important factors that would impact  the results of our study. 

In particular, spherical winds driven by radiative pressure could dramatically alter the flow of the cold filamentary accretion once the BH has  become sufficiently massive and may -- at least in principle -- prevent the stellar bulge from becoming  as tightly bound and massive as we found in our simulations.
A simple calculation of the potential effect of AGN feedback on the binding energy of the bulge component shows that a $M_{\rm BH}=10^9 M_\odot$ BH has enough energy $E_{\rm AGN}=\eta \times 0.1 M_{\rm BH}c^2=\eta \times 2\cdot 10^{62}$ ergs to efficiently remove all baryons from the bulge with a total energy of a few $10^{59}$ ergs, even considering low conversion efficiency of AGN luminosity into mechanical energy output (e.g. $\eta=0.01$).
It is, however, not obvious what level of feedback would be needed to stop the accretion flow, because observed AGN feedback is often focused into a small/moderate solid angle (blazars, jets) and may therefore not significantly impact the cold filamentary accretion or the capture of star-forming clumps (as illustrated in fig.~\ref{fig:clump_migration}).
It is also not obvious to what extent feedback schemes currently implemented in numerical simulations actually properly capture the coupling of the released energy and momentum to the interstellar and circumgalactic media. 
This problem will thus not be straightforward to address, and we postpone this issue to a future study of the impact of AGN feedback on the gas inflow onto the central BH by testing different modes of AGN feedback (collimated jets or spherical winds following the \citealp{duboisetal12} model).
The detailed mode and efficiency of star formation will also strongly affect how compact the bulge will become~\citep{agertzetal11} and, more importantly, how much gas will be left over for the feeding of the central supermassive BH. 

Last but not least, there are still numerical issues with the sinking of cold gas in hot gaseous atmospheres, 
which in the past have  led to the formation of overly prominent galactic bulges.  This appears to be, however,  
a more serious problem  for  SPH codes  than for AMR-based hydrodynamical codes like {\sc ramses}~\citep{torreyetal11,scannapiecoetal11}. The fact that {\sc ramses} simulations 
of more common halos at lower redshifts with a very similar setup to ours result in prominent extended disc 
and moderately dominant and less compact bulges provides some reassurance here. 
We hope to be able to report progress on  some of these issues in a forthcoming study. 

\subsection{Implications for the formation/build-up of compact bulges and their central supermassive BHs }
\label{sec:Implications}

The most striking result of our simulations is the large fraction of gas that settles into a very compact bulge. 
In fact the  circular velocity of the bulge significantly exceeds the circular velocity of the host halo
and with a circular velocity of 900 km/s is significantly larger than that of even the most massive galactic bulges 
at low redshift. The absence of such strongly bound galactic bulges in the nearby Universe  is not necessarily an observational concern. As we have discussed earlier, only about a ten thousand or less of such objects are expected to have formed in the whole Universe. Furthermore, such an ultra-compact central region would not be easy to detect in a much larger 
and more massive galactic bulge surrounding it. It may in fact be mistaken as the direct signature of a central 
BH. A strong evolution of the characteristic  sizes of galactic bulges is actually observed. 
Galactic bulges  appear to be  about a factor 3-5 smaller  at $z\sim2-3$~\citep{daddietal05, cimattietal08, vandokkumetal08, whitakeretal12, oseretal12}
than they are at $z=0$.  
Measuring stellar  velocity dispersions at these redshifts is very difficult and expensive in terms of telescope time but there is the intriguing  case described by~\cite{vandokkumetal09} with  a claimed velocity dispersion of 510 km/s at $z= 2.1$. 
  At the redshifts we are considering here ($z\sim6$) the only  observational data probing  the depth of the potential wells of the galaxies hosting the billion solar mass BHs are a few CO measurements with rather low velocity width~\citep{wangetal10}, but it is far from clear to what extent this actually probes the depth of the central potential well  of the galactic bulge presumably hosting these supermassive BHs.

 With the resolution of our simulation there is still some way to go to reach the sphere of influence 
 even of a billion solar mass BH  but there is definitely enough low angular momentum gas funneled to the centre  to build-up the masses reported for the most luminous QSOs at $z\sim 6 - 7$.
 In fact if we take the the circular velocity of the galactic bulges in our simulations at face value, a non-evolving  $M_{\rm BH}/\sigma$ -relation would predict BHs well in excess of ten billion solar masses.
 More important is that the gas arrives at the centre continuously as required to  explain  the sustained gas supply at close to the Eddington rate over cosmological timescales  which has also  been reported by \cite{sijackietal09} and \cite{dimatteoetal12}  from their simulations. As we have discussed, a significant fraction of this gas  streams  straight into the bulge without ever properly settling into a disk.  In fact a fraction of the gas actually streams  straight into the centre of the bulge as defined by the resolution limit of our simulations.  This gas spirals inwards in about two dynamical times from a tenth of the virial radius 
and while  the streams carrying the gas into the halo  are  aligned  in  a preferred plane at large radii, at small radii the inflow is largely coming from random directions in good agreement with the lack of correlation of the observed directions of accretion driven jets with the orientation of galaxies \citep{hopkinsetal11, king&pringle07}.
The largest  remaining uncertainty in this regard  is  the competition between accretion onto the central supermassive BH and consumption of gas by star formation as well as the efficiency of stellar and AGN feedback which we have not addressed here. 
However, measured accretion rates onto the bulge ($200\, \rm M_\odot/yr$) are largely sufficient to continuously supply gas for a BH accreting at its Eddington limit or above ($\dot M_{\rm Edd}=22 \, \rm M_\odot/yr$ for a $10^9\, \rm M_\odot$ BH).
Bulges as  strongly gravitationally bound as those in our simulation here would, however, ensure that even extreme feedback would have a hard time to revert the continuous inflow of gas. 
It  may indeed take supermassive BHs with several billion solar masses to do this and limit further growth as discussed by \cite{sijackietal09} and \cite{dimatteoetal12}. 
We leave a further investigation of these issues to future work. 

 Note that the picture presented here offers an attractive explanation for the apparent change 
in the nature of build up of supermassive BHs and galactic bulges with redshift
(see  \citealp{haehnelt&rees93} for some early ideas).  At high redshift rare   halos are continuously fed~\citep{shankaretal11}
with low angular momentum gas  streaming in along the filaments penetrating 
straight to the compact bulge.  At intermediate redshift  (say $ z\sim 2-4$) 
and probably also in less rare halos at high redshift  the gas is fed into the halos with progressively  more angular momentum and hangs around more and more  in a disc which is intermittently  gravitationally unstable. The role of major (and minor) mergers becomes more important but not necessarily dominant.  The duty cycle  of AGN powered by accretion onto supermassive  BHs  decreases.   At low redshift it becomes harder and harder for the cold filamentary  streams   to penetrate  deep  into the halos~\citep{dekel&birnboim06}, many especially massive galaxies 
are now incorporated in galaxy groups and clusters where their direct connection  to the  the cosmic web
is interrupted  and this mode of feeding AGNs  ceases.  The cold gas still present at low redshift now sits preferentially in lower mass halos (see ~\citealp{kauffmann&heckman09} for an extensive discussion of the fueling of AGN at low redshift).  The gas in these lower mass halos  has  high angular momentum which it  has to lose   by secular processes to reach the central BH.  The more massive halos can support a hot 
atmosphere  at low redshift (or have been incorporated into a galaxy group or cluster).  Continuous slow accretion of hot gas through cooling flows starts therefore to dominate instead the feeding of the BHs  in the massive halos at low redshift~\citep{duboisetal10, duboisetal11}.

\section{Conclusions}
\label{sec:conclusion}

We have investigated here the formation of a compact central bulge in two rare very massive halos 
at redshift $z=6$.  Our main results are as follows.

\begin{itemize}

\item{The typical path for a parcel of fluid entering a halo and eventually ending up in the bulge, is to stream out of 
a void into the surrounding wall, from there into the filaments defining the intersection of the cosmic walls/sheets, and then 
along the filaments which penetrate deep into the halo.}

\item{ Early on, most  of the  baryons ending up in the bulge
stream nearly radially along the cold filamentary infall directly into a very compact bulge. As time progresses an increasing fraction of the baryons settles first into a compact gravitationally unstable transient central disc from where they  progress rapidly into the bulge. By redshift $z=6$ about half the baryons forming the bulge move first through this surrounding disc before reaching the bulge.}

\item{In the outer part of the halo  the angular distribution of the baryonic inflowing matter is stable and reflects the angular distribution of the cosmic web, while well inside the virial radius the angular distribution becomes much more random and changes rapidly.}

\item{A small fraction ($\sim$ 10 \%) of the baryons in the halos falls directly from the structures of the cosmic web into the centre of the halos in the form of proto-galactic clumps; these clumps are carried along the filamentary streams, that represent more than half of the total accretion of gas, and join the compact central bulge as minor mergers. }   

\item{In our simulations most of the baryons in the bulge turn into stars with a remaining gas fraction of 5-30\%. The gas fraction  is, however,  strongly  dependent on the resolution  and the star formation criterion  of the simulation. The gas fraction should also be strongly affected by feedback both from  the stars but even more by feedback from the accreting central supermassive BH once the BH  has become sufficiently massive.  Note again that
the simulations used here  do  not include any feedback.}

\item{The baryon dominated bulge has angular momentum  a factor 10-30 lower than the average specific angular momentum of the  baryons in the two simulated halos. The baryonic material which is mainly in the form of stars is  extremely strongly bound and compact with a peak of the circular velocity approaching 1000 km/s at a radius $< 100$pc. }

\item{The low value of the angular momentum is due to three effects. First the bulge is preferentially made from material which already has a somewhat lower specific angular momentum than average as it enters the halo along the  filaments of the cosmic web. Second the baryonic matter entering the halo from the cosmic environment does so with  a broad angular momentum distribution. 
As the gas streams in cold the lower angular momentum material falls quickly to the centre of the halo due to the lack of pressure support where the randomly oriented angular momentum  effectively cancels. 
Third there is additional angular momentum loss due to dynamical friction on inspiralling proto-galactic clumps, exchange between spin and orbital angular momentum and 
gravitational instabilities in the disc.}     

\item{The fraction of gas  streaming  nearly radially to the centre of the halo 
making up the low-angular momentum bulge appears to strongly increase for rare halos. 
We argue that  this is due to the fact that isolated rare halos are connected to the cosmic web more 
isotropically and with a larger  number of filaments.  This leads both to lower specific angular momentum when the baryonic matter enters the halo and to more complete angular momentum cancellation when the low angular momentum material streams cold  into the  compact bulge. 
}

\end{itemize}

Numerical simulations of galaxy formation are still subject to a  range of uncertainties both numerical and 
in terms of the implemented physics in particularly with respect to feedback due to stars and AGN. The filamentary 
cold inflow of gas appears nevertheless  to emerge as the key ingredient in the formation of the observed compact discs and bulges at high redshift. As we have discussed here it may also be key to  the efficient  build up of the several billion solar mass black holes inferred to exist as early  as $z\sim 6-7$.  


\section*{Acknowledgments}
We are grateful to J. Silk, R. Teyssier  and J. Binney for helpful comments and thank A. Dekel for a stimulating talk on related matters given at IAP. We would like to thank the anonymous referee for a very constructive report that helped to improve the quality of the paper. 
YD is supported by an ERC advanced grant (DARK MATTERS).
The simulations presented here were run on the DiRAC facility jointly funded by STFC, the Large Facilities Capital Fund of BIS and the University of Oxford.
This research is part of the Horizon-UK project.
Let us thank D.~Munro for freely distributing his {\sc \small  Yorick} programming language and opengl interface (available at {\tt http://yorick.sourceforge.net/}).
We thank the Alliance program (EGIDE), J. Bergeron and the ``programme visiteur" of the institut d'Astrophysique de Paris for funding.
 We also acknowledge support from the Franco-Korean PHC STAR program  and the France Canada Research Fund.

\bibliographystyle{mn2e}
\bibliography{author}

\appendix

\section{Tracer particles to follow the Lagrangian flow of gas}
\label{App:tracer}

\begin{figure}
   \includegraphics[width=1.0\columnwidth]{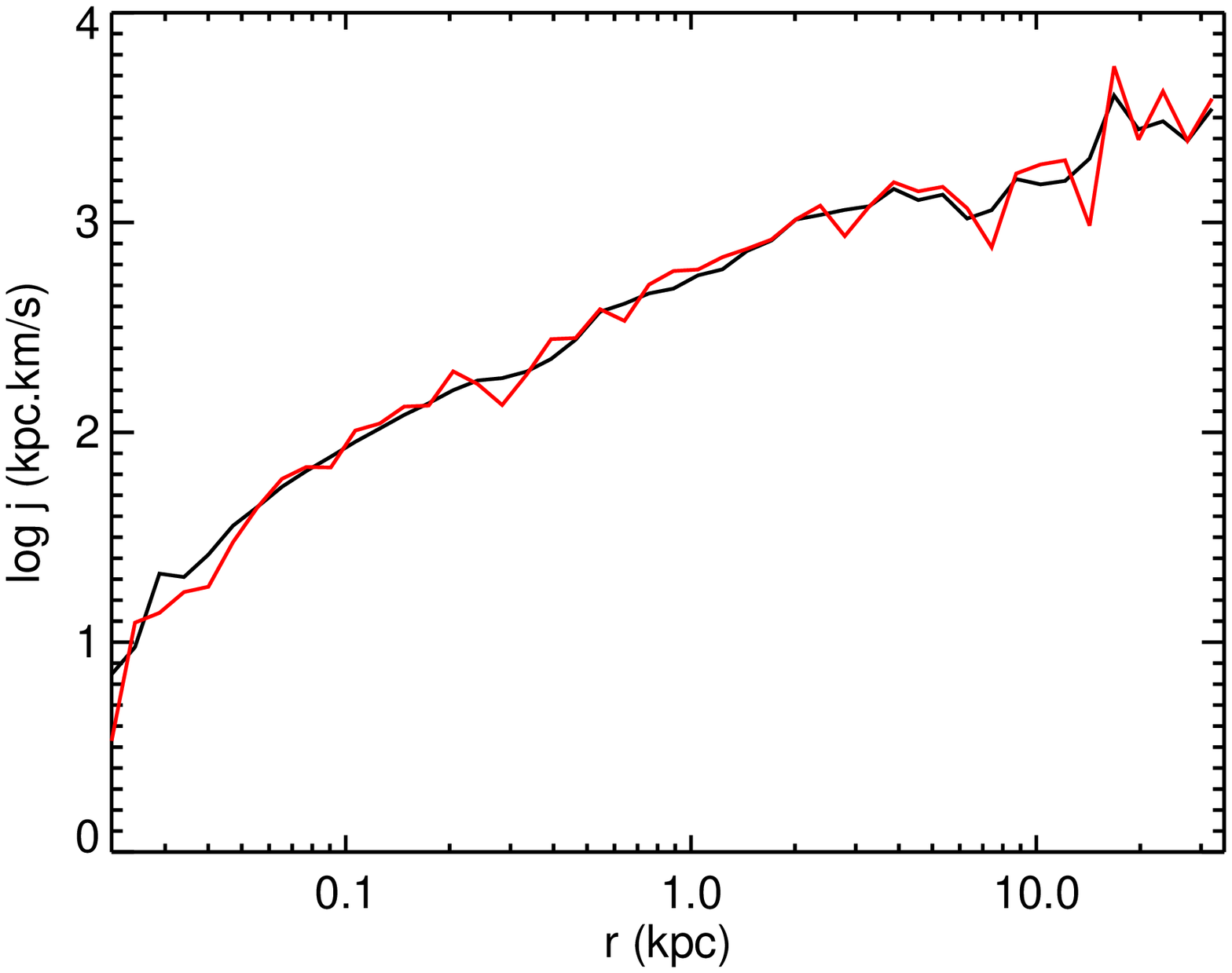}
      \caption{Gas specific angular momentum computed on the actual AMR grid (solid line), or using tracer particles (dashed line) for the SHhr halo at $z=6$. Profiles are in a very good agreement.}
\label{fig:angmomtracer_comp}
\end{figure}

Tracer particles can be used in Eulerian techniques to follow the Lagrangian path of gas elements.
They are passively advected with the gas flow using a simple forward Eulerian step
\begin{equation}
x_{i}^{n+1}=x_i+v_i^n \cdot \Delta t^n\,,
\end{equation}
using the surrounding gas velocity $v_i^n$ at time $t^n$ from the $x_i^n$ position of the i-th tracer particle.
The gas velocity is interpolated using a CIC interpolation scheme on the grid with the size of the cloud equal to the size of host cell.
It is a crude, but simple, interpolation of the true gas flow determined by the Euler scheme.
More precise (and more complicated) updates of tracer positions could be used (such as Runge-Kutta updates) using velocity of the gas at different times.
We evaluated the agreement between the gas and tracer particles specific angular momentum profiles in the SHhr simulation at $z=6$ and find very good agreement between the two (fig.~\ref{fig:angmomtracer_comp}).

\end{document}